%% file: fse21-final.tex
\PassOptionsToPackage{table,xcdraw}{xcolor}
\documentclass[sigconf,screen]{acmart}

\usepackage{booktabs}   %% For formal tables:
\usepackage{array}
\usepackage{amsmath,amsfonts}
\usepackage{algorithm}
\usepackage[noend]{algpseudocode}
\usepackage{graphicx}
\usepackage{textcomp}
\usepackage{float} 
\usepackage{listings}
\usepackage{xspace}
\usepackage{amsthm}
\usepackage{subfig}
\usepackage{multirow}
\usepackage{url}
\usepackage{balance}

\usepackage{tikz}
\usetikzlibrary{shapes}

\DeclareMathOperator{\EX}{\mathbb{E}}% expected value

\usepackage[skins]{tcolorbox}

\newtcolorbox{myframe}[2][]{%
  enhanced,colback=white,colframe=black,coltitle=black,
  sharp corners,
  toprule=1.0pt,
  rightrule=0.3pt,
  leftrule=0pt,
  bottomrule=0pt,
  fonttitle=\itshape\scshape\large,
  left=0pt,right=5pt,top=5pt,bottom=3pt,
  attach boxed title to top right={yshift=-0.3\baselineskip-0.4pt,xshift=-5mm},
  boxed title style={tile,size=minimal,left=0.2mm,right=0.5mm,
    colback=white,before upper=\strut},
  title=#2,#1
}

\newcommand*\circled[1]{\tikz[baseline=(char.base)]{
    \node[shape=circle,draw,inner sep=2pt] (char) {#1};}}

\newcommand{\tool}{\textsc{IVDetect}\xspace}

\newcolumntype{L}[1]{>{\raggedright\arraybackslash}p{#1}}

\newcommand{\code}[1]{{\small\textsf{#1}}}
\usepackage{amsthm}
 \definecolor{dkgreen}{rgb}{0,0.6,0}
\definecolor{gray}{rgb}{0.5,0.5,0.5}
\definecolor{mauve}{rgb}{0.58,0,0.82}
\AtBeginDocument{%
  \providecommand\BibTeX{{%
    \normalfont B\kern-0.5em{\scshape i\kern-0.25em b}\kern-0.8em\TeX}}}

\setcopyright{acmcopyright}
\acmPrice{15.00}
\acmDOI{10.1145/3468264.3468597}
\acmYear{2021}
\copyrightyear{2021}
\acmSubmissionID{fse21main-p549-p}
\acmISBN{978-1-4503-8562-6/21/08}
\acmConference[ESEC/FSE '21]{Proceedings of the 29th ACM Joint European Software Engineering Conference and Symposium on the Foundations of Software Engineering}{August 23--28, 2021}{Athens, Greece}
\acmBooktitle{Proceedings of the 29th ACM Joint European Software Engineering Conference and Symposium on the Foundations of Software Engineering (ESEC/FSE '21), August 23--28, 2021, Athens, Greece}

\begin{document}

%\title{The Name of the Title is Hope}

\title[Vulnerability Detection with Fine-Grained Interpretations]{Vulnerability Detection with Fine-Grained Interpretations}

\author{Yi Li}
\affiliation{
	\institution{New Jersey Inst. of Technology}
%	\city{Newark}
	\state{New Jersey}
	\country{USA}
}
\email{yl622@njit.edu}
\author{Shaohua Wang}
\authornote{Corresponding Author}
\affiliation{
	\institution{New Jersey Inst. of Technology}
%	\city{Newark}
	\state{New Jersey}
	\country{USA}
}
\email{davidsw@njit.edu}
\author{Tien N. Nguyen}
\affiliation{
	\institution{University of Texas at Dallas}
%	\city{Richardson}
	\state{Texas}
	\country{USA}
}
\email{tien.n.nguyen@utdallas.edu}

\renewcommand{\shortauthors}{Li, Wang, and Nguyen}

%\begin{abstract}
%  A clear and well-documented \LaTeX\ document is presented as an
%  article formatted for publication by ACM in a conference proceedings
%  or journal publication. Based on the ``acmart'' document class, this
%  article presents and explains many of the common variations, as well
%  as many of the formatting elements an author may use in the
%  preparation of the documentation of their work.
%\end{abstract}

\input{sections/abstract}





\maketitle

\input{sections/intro}

\input{sections/motiv}
\input{sections/appro}

\input{sections/exper}

\input{sections/threats}
\input{sections/relat}

\input{sections/concl}

\section*{Acknowledgments}
This work was supported in part by the US National Science Foundation
(NSF) grants CCF-1723215, CCF-1723432, TWC-1723198, CCF-1518897, and
CNS-1513263.

%\bibliographystyle{ACM-Reference-Format}
%\bibliography{sample-base}

\balance

\bibliographystyle{ACM-Reference-Format}

\bibliography{sections/FL,sections/icse21IntVD,sections/Reference}

\end{document}

%% file: sections/abstract.tex
\begin{abstract}

Despite the successes of machine learning (ML) and deep learning (DL)
based vulnerability detectors (VD), they are limited to providing only
the decision on whether a given code is vulnerable or not, without
details on {\em what part of the code is relevant to the detected
  vulnerability}.
We present {\tool}, an {\em interpretable vulnerability detector} with
the philosophy of using Artificial Intelligence (AI) to detect
vulnerabilities, while using Intelligence Assistant (IA) via providing
VD interpretations
%at the {\em fine-grained level}
in terms of vulnerable statements.

For vulnerability detection, we separately consider the vulnerable
statements and their surrounding contexts via data and control
dependencies. This allows our model better discriminate vulnerable
statements than using the mixture of vulnerable code and~contextual
code as in existing approaches.
In addition to the coarse-grained vulnerability detection result, we
leverage {\em interpretable AI} to provide users with {\em
  fine-grained interpretations that include the sub-graph in the Program Dependency Graph (PDG) with the crucial statements} that are relevant to the detected
vulnerability.
%
%The interpretations include the sub-graph in the PDG of the given code
%including crucial statements, and variables with program slices, and
%the Common Weakness Enumeration (CWE) that are relevant to the
%detected vulnerability.
%
%First, we model the VD via Graph Convolution Network (GCN) with
%feature-attention (FA) mechanism. With the decision and the trained
%FA-GCN model, we develop a graph-based interpretation model via
%GNNExplainer with the formulation of finding the minimal PDG sub-graph
%and the minimal set of features that minimizes the prediction scores
%between using the entire graph or entire set of features and the
%minimal ones.
%
Our empirical evaluation on vulnerability databases shows that {\tool}
outperforms the existing DL-based approaches by 43\%--84\% and
105\%--255\% in top-10 nDCG and MAP ranking scores.
{\tool} correctly points out the vulnerable statements relevant to the
vulnerability via its interpretation~in 67\% of the cases with a top-5
ranked list. It improves over baseline interpretation models by
12.3\%--400\% and 9\%--400\% in accuracy.

%also reduces {\bf 80\%} of the numbers of statements that need to be
%investigated.

\end{abstract}

%% file: sections/intro.tex
\section{Introduction}
\label{intro}

Software vulnerabilities have caused substantial damage to society's software infrastructures. 
Automated vulnerability detection
(VD) approaches can be broadly classified into two categories: program
analysis (PA)-based~\cite{FlawFinder, RATS, viega2000its4, Checkmarx,
  HPFortify, Coverity} and machine learning
(ML)-based~\cite{scandariato2014predicting,neuhaus2007predicting,shin2010evaluating}.
The PA-based VD techniques have often focused on solving the {\em
  specific types of vulnerabilities} such as
BufferOverflow~\cite{BufferOverFlow}, SQL Injection~\cite{SQLInj},
Cross-site Scripting~\cite{Cross-siteScripting}, Authentication
Bypass~\cite{AuthBypassSpoofing}, etc. In addition to those types, the
more general software vulnerabilities, e.g., in API usages of
libraries/frameworks, have manifested in various forms. To detect
them, machine learning (ML) and deep learning (DL) have been leveraged
to implicitly learn the patterns of vulnerabilities from prior
vulnerable
code~\cite{li2018vuldeepecker,zhou2019devign,harer2018automated}.

Despite several advantages, the ML/DL-based VD approaches are still limited to providing only coarse-grained detection results~on whether an entire given method is vulnerable or not. 
In comparison with the
PA-based approaches, they fall short in the ability to elaborate on
the {\em fine-grained details} of the lines of code with specific
statements that might be involved in the detected vulnerability. One
could use fault localization (FL) techniques~\cite{keller2017critical}
to locate the vulnerable statements, however they require
large, effective test suites. Due to such feedback at the coarse
granularity from the existing ML/DL-based VD tools, developers would
not know where and what to look for and to fix the vulnerability in
their code. This hinders them in investigating the potential
vulnerabilities.

%no need
%decide to continue and waste time/effort if it is NV

%Tien: AI versus AI
%Artificial Intelligence (AI) to detect the vulnerable method, while it
%uses Intelligence Assistant (IA) with interpretations to support
%developers in further investigation.

%providing more would save time and effort in investigation

To raise the level of ML/DL-based VD, we present {\tool}, an {\em
  interpretable VD} with the philosophy of using {\em Artificial
  Intelligence} to detect coarse-grained vulnerability, while
leveraging {\em Intelligence Assistant} via interpretable ML to provide
fine-grained interpretations in term of vulnerable statements relevant to the
vulnerability.

%providing VD at  the fine-grained level via interpretations  of the VD
%to support users in further investigation.

For coarse-grained {\em vulnerability detection}, our novelty is the
{\em context-aware representation learning of the vulnerable code}.
During training, the existing ML/DL-based VD
approaches~\cite{li2018vuldeepecker,zhou2019devign} take the entire
vulnerable code in a method as the input without distinguishing the
vulnerable statements from the surrounding contextual code. Such
distinction from vulnerable code and the contexts during training
enable {\tool} to better learn to discriminate the vulnerable code and
benign ones. We represent source code via program dependence graph
(PDG) and we treat the vulnerability detection problem as graph-based
classification via Graph Convolution Network (GCN)~\cite{GCN16} with
feature-attention (FA), namely FA-GCN.
%
%A node represents a statement, and an edge is for a data or control
%dependency among nodes. The features of a node model the properties of
%a program element or a statement.
The vulnerable statements, along with surrounding code, are encoded
during the code representation learning.
%
%The label function $f$ maps a method in the training dataset to a
%set of two elements: vulnerable ($\mathcal{V}$) or non-vulnerable
%($\mathcal{NV}$).

For {\em fine-grained interpretation}, as the given method is deemed
as vulnerable by {\tool}, our novelty is to {\em leverage
  interpretable ML~\cite{GNNExplainer} to provide the interpretation
  in term of the vulnerable statements as part of the PDG that are
involved to the detected vulnerability}.  The rationale for choosing
PDG sub-graph as an interpretation is that a vulnerability often
involves the data and control dependencies among the
statements~\cite{pham2010detection}.

To derive the vulnerable statements as the interpretation, we leverage
the interpretable ML model, GNNExplainer~\cite{GNNExplainer}, that
{\em ``explains'' on why a model has arrived at its decision}.
Specifically, after vulnerability detection, to produce
interpretation, {\tool} takes as input the FA-GCN model along with
its decision (vulnerable or not), and the input~PDG $G_M$ of the given
method $M$. The goal is to find the interpretation subgraph, which is
defined as a minimal sub-graph $\mathcal{G}$ in the PDG of $M$ that
{\em minimizes the prediction scores between using the entire $G_M$
  and using $\mathcal{G}$}.
%Similar treatment is applied to find a minimal set $\mathcal{X}$ of
%the feature set $X_M$.
To that end, we leverage GNNExplainer~\cite{GNNExplainer} in which the
searching for $\mathcal{G}$ is formulated as the learning of
the edge-mask set $EM$.
%and the feature-mask set $FM$.
The idea is that if an edge belongs $EM$, i.e., {\em if it is removed from
$G_M$, and the decision of the model is affected, then the edge is
crucial and must be included in the interpretation for the detection
result}. Thus, the minimal sub-graph $\mathcal{G}$ in PDG contains the
nodes and edges, i.e., the {\em crucial statements and program
dependencies, that are most decisive/relevant to the detected
vulnerability} when the decision is vulnerable.
%Thus, if the decision is vulnerable, the minimal sub-graph
%$\mathcal{G}$ in PDG represents the crucial statements and
%dependencies that are most decisive to the detected vulnerability.
%
%Similarly, the minimal subset $\mathcal{X}$ of the feature set $X_M$
%represents the minimal set of crucial variables decisive/relevant to
%the detected~vulnerability.

%The backward program slices are also computed.

%We perform an extensive empirical evaluation on
%{\tool}.

Using our results, a practitioner would 1) examine the ranked list of potentially vulnerable methods, and 2) use the interpretation to further investigate what statements in the code that caused the model to predict that vulnerability.

We conducted several experiments to evaluate {\tool} in both
vulnerability detection at the method level and interpretation in term
of vulnerable statements. We use 3 large C/C++ vulnerability datasets:
Fan~\cite{fan2020msr}, Reveal~\cite{chakraborty2020deep} and
FFMPeg+Qemu~\cite{zhou2019devign}.
For the method-level VD, our results show that {\tool} outperforms the existing ML/DL-based approaches~\cite{li2018vuldeepecker, zhou2019devign, li2018sysevr, russell2018automated,chakraborty2020deep} by 43\%--84\% and 105\%--255\% at the top 10
list for two ranking scores nDCG and MAP, respectively. For the
statement-level interpretation, {\tool} correctly points out the
vulnerable statements relevant to the vulnerability in 67\% of the
cases with a top-5 ranked list. It improves over the baseline ATT~\cite{GNNExplainer} and
		GRAD~\cite{GNNExplainer} interpretation models by 12.3\%--400\% and 9\%--400\% in accuracy,
respectively.

%
%Our results show that {\tool} can correctly generate the relevant
%sub-graphs as interpretation for the predictions of a graph-based
%detection model for 73.8\% of the vulnerable cases, and 67.3\% of the
%non-vulnerable cases. It helps reduce ~80\% of the number of
%statements that needs to be inspected. In addition, in 76.2\% of the
%cases, it is able to identify the vulnerability-relevant key variables
%in the top-5 ranked variables in the context that on average a method
%has 15 variables in the dataset. Moreover, it can retrieve the
%relevant CWEs for the sub-graphs with a top-1 accuracy of
%68.2\%. Also, our graph-based vulnerability detection model can
%generate comparable and complementary results to the state-of-the-art
%DL-based vulnerability detection tools.
%{\it Our data is available at~\cite{Interpretation2021icse}.}

The contributions of this paper include:

\noindent {\bf \underline{A. Interpretable VD with Fine-grained Interpretations}}

{\bf a. Vulnerability Detection with Fine-grained Interpretations}:
{\tool} is the first approach to leverage interpretable ML to enhance
VD with {\em fine-grained} details on PDG sub-graphs, statements, and
dependencies relevant to the detected vulnerability.

{\bf b. Context-aware Representation Learning} of vulnerable code: The
novelty of our representation learning of vulnerable code is {\em our
  consideration of the contextual code surrounding the vulnerable
  statements and fixes} to better train the VD model.

\noindent {\bf \underline {B. Empirical Evaluation.}} Our results show
          {\tool}'s high accuracy in both detection and interpretation
          (See data/results at~\cite{Interpretation2021icse}).

%% file: sections/motiv.tex
\vspace{-8pt}
\section{Motivation}

\input{sections/motiv-example}

\input{sections/motiv-keyideas}

%% file: sections/motiv-example.tex
\subsection{Motivating Example}
\label{motiv:sec}

Figure~\ref{fig:motiv_1} shows the method
\code{ec$\_$device$\_$ioctl$\_$xcmd} in Linux 4.6, which constructs
the I/O control command for the CromeOS devices. This is listed as a
vulnerable code within Common Vulnerabilities and Exposures
(CVE-2016-6156) in the National Vulnerability Database.
%(NVD).

%{\em ``Race condition in the ec\_device\_ioctl\_xcmd function in
%drivers/platform/chrome/cros\_ec\_dev.c in the Linux kernel before 4.7
%allows local users to cause a denial of service (out-of-bounds array
%access) by changing a certain size value, aka a "double fetch"
%vulnerability.''}

The commit log of the corresponding fix stated that 

{\em ``At line 6
  and line 13, the driver fetches user space data by pointer
  \code{arg} via \code{copy$\_$from$\_$user()}. The first fetched
  value (stored in \code{u$\_$cmd}) (line 6) is used to get the
  \code{in$\_$size} and \code{out$\_$size} elements and allocation a
  buffer (\code{s$\_$cmd}) at line 10 so as to copy the whole message
  to driver later~at~line 13, which means the copy size of the whole
  message (\code{s$\_$cmd}) is based on the old value
  (\code{u$\_$cmd.outsize}) from the first fetch. Besides, the whole
  message copied at the second fetch also contains the elements of
  \code{in$\_$size} and \code{out$\_$size}, which are the new
  values. The new values from the second fetch might be changed by
  another user thread under race~condition, which will result in a
  double-fetch bug when the inconsistent values are used.''}  

Thus, to
fix this bug, a developer added the code at lines 17--21 to make sure
that \code{u$\_$cmd.outsize} and \code{u$\_$cmd.insize} have not
changed due to race condition between the two fetching
calls. Moreover, memory access might be also beyond the array
boundary, causing a buffer overflow within the method call
\code{cros$\_$ec$\_$cmd$\_$xfer(...)}, when the command is transferred
to the ChromeOS device at \underline{line 23}.

Another issue is at \underline{line 27} with
\code{copy$\_$to$\_$user}. The method call \code{cros$\_$ec$\_$cmd}
\code{$\_$xfer(...)} can set \code{s$\_$cmd-$>$insize} to a lower
value. Thus, the new smaller value must be used to avoid copying too
much data~to the user: \code{u$\_$cmd.insize} at line 27 is changed
into \code{s$\_$cmd-$>$insize}.

\begin{figure}[t]
	\centering
	\renewcommand{\lstlistingname}{Method}
	\lstset{
		numbers=left,
		numberstyle= \tiny,
		keywordstyle= \color{blue!70},
		commentstyle= \color{red!50!green!50!blue!50},
		frame=shadowbox,
		rulesepcolor= \color{red!20!green!20!blue!20} ,
		xleftmargin=2em,xrightmargin=1em, aboveskip=1em,
		framexleftmargin=0.5em,
		language=Java,
		basicstyle=\tiny\ttfamily,%scriptsize\ttfamily,
		moredelim=**[is][\color{red}]{@}{@},
		escapeinside= {(*@}{@*)},
}
	\begin{lstlisting}
static long ec_device_ioctl_xcmd(struct cros_ec_dev *ec, void __user *arg)
{
	long ret;
	struct cros_ec_command u_cmd;
	struct cros_ec_command *s_cmd;
	if (copy_from_user(&u_cmd, arg, sizeof(u_cmd)))
		return -EFAULT;
	if ((u_cmd.outsize > EC_MAX_MSG_BYTES) || (u_cmd.insize > EC_MAX_MSG_BYTES))
		return -EINVAL;
	s_cmd = kmalloc(sizeof(*s_cmd) + max(u_cmd.outsize, u_cmd.insize), GFP_KERNEL);
	if (!s_cmd)
		return -ENOMEM;
    if (copy_from_user(s_cmd, arg, sizeof(*s_cmd) + u_cmd.outsize)) {
		ret = -EFAULT;
		goto exit;
	}
+	if (u_cmd.outsize != s_cmd->outsize ||
+		u_cmd.insize != s_cmd->insize) {
+		ret = -EINVAL;
+		goto exit;
+	}
	s_cmd->command += ec->cmd_offset;
	@ret = cros_ec_cmd_xfer(ec->ec_dev, s_cmd);@
	/* Only copy data to userland if data was received. */
	if (ret < 0)
		goto exit;
-	@if (copy_to_user(arg, s_cmd, sizeof(*s_cmd) + u_cmd.insize))@
+	if (copy_to_user(arg, s_cmd, sizeof(*s_cmd) + s_cmd->insize))
		ret = -EFAULT;
exit:
	kfree(s_cmd);
	return ret;
}
	\end{lstlisting}
	\vspace{-0.16in}
	\caption{CVE-2016-6156 Vulnerability in Linux 4.6}
        \vspace{-0.06in}
	\label{fig:motiv_1}
%	\vspace{-5pt}
\end{figure}

This vulnerable code could potentially cause the damages such as
denial of service, buffer overflow, program crash, etc.
%Recent advances in
Deep learning (DL) advances enable several
approaches~\cite{zhou2019devign,li2018vuldeepecker} to {\em implicitly
  learn} from the history the patterns of vulnerable code, and to
detect {\em more general vulnerabilities}.
%in a new given code.
However, they are still limited in comparison with program
analysis-based approaches in the ability to provide any detail on the
{\em fine-grained} level of the vulnerable statements,
%involving in a vulnerability,
and on why the model has decided on the vulnerability.
For example, the PA-based approaches, e.g., a race detection
technique could potentially detect the involvement of the two fetching
statements at line 6 and line 13. The method in
Figure~\ref{fig:motiv_1} might be deemed as vulnerable by a DL-based
model. But without any fine-grained details, a developer would not
know where and what to investigate next. This would make the output of
a DL model less constructive in VD. Moreover, a fault localization
technique~\cite{keller2017critical}, which locates buggy statements,
would need a large, effective test suite.

%does not work in this case because it requires a large test suite.

%whether a developer would be willing to investigate
%further the code.

Regarding detection, the existing DL-based
approaches~\cite{zhou2019devign,li2018vuldeepecker} do not fully
exploit all the available information on the vulnerable code during
training. For example, during training, we know that lines 23 and 27
are vulnerable/buggy, and other {\em relevant statements via
  data/control dependencies provide contextual information for the
  vulnerable ones}. However, the existing
approaches~\cite{zhou2019devign,li2018vuldeepecker} do not consider
the vulnerable statements and do not use the contextual code to help a
model discriminate the vulnerable and non-vulnerable ones. The entire
method would be fed to a DL model.

%in Figure~\ref{fig:motiv_1}

%\subsection{{\tool} Approach}

\vspace{-0.06in}
\subsubsection*{\bf {\tool} Approach}
\begin{figure}[t]
  \centering \includegraphics[width=3.3in]{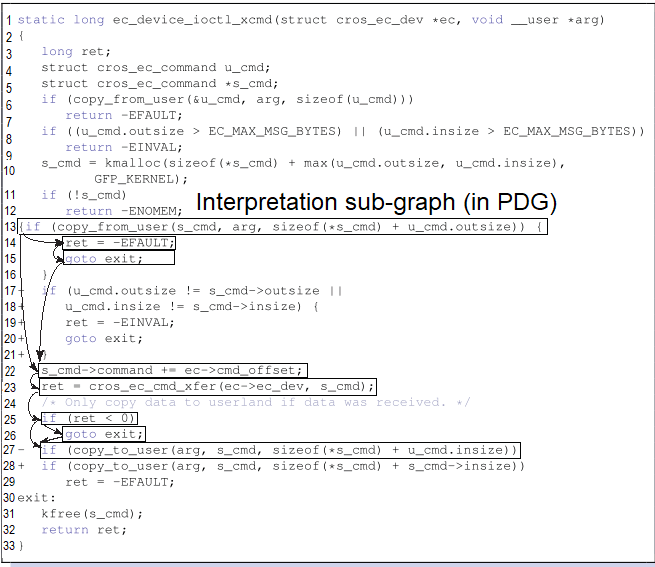}
  \vspace{-0.06in}
	\caption{Interpretation Sub-Graph for Figure~\ref{fig:motiv_1}}
%        \vspace{-0.16in}
	\label{fig:pdg}	
	\vspace{-0.06in}
\end{figure}

We introduce {\tool}, an DL-based, {\em interpretable vulnerability
  detection} approach that goes beyond the decision of vulnerability
by providing the {\em fine-grained} interpretation in term of the
vulnerable statements. Specifically, as the method is deemed as
vulnerable by {\tool}, it will provide a {\em list of important
  statements as part of the program dependence graph (PDG) that are
  relevant to the detected vulnerability}. For example, it provides
the partial sub-graph of the PDG including the statements at the lines
13--15, 22--23, and 25--27 in Figure~\ref{fig:pdg} for the vulnerable
code at line 23 and line 27. We use the PDG sub-graph including
important statements for fine-grained VD since they will give a
developer the hints on the program dependencies relevant to the vulnerability
for further investigation.  Moreover, if our model determines the code
as~non-vulnerable, it can also produce the key sub-graph of the PDG
with key statements that are deemed to be safe.

%Tien

%% file: sections/motiv-keyideas.tex
\subsection{Key Ideas and Architecture Overview}

\begin{figure}[t]
	\centering
	\includegraphics[width=3.3in]{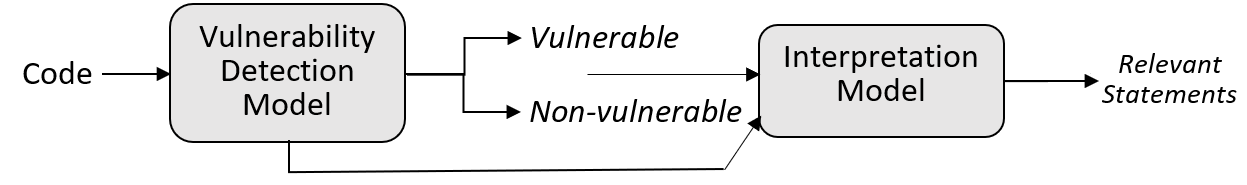}
        \vspace{-0.06in}
	\caption{Overview of {\tool}}
	\label{fig:arch}
	\vspace{-0.06in}	
\end{figure}

%Our philosophy is that we use {\em Artificial Intelligence (AI) to
%  detect the vulnerable method, while using Intelligence Assistant
%  (IA) with fine-grained VD interpretations to assist developers in
%  further investigation}.
{\tool} has two main modules (Figure~\ref{fig:arch}): graph-based
vulnerability detection model, and graph-based interpretation
model. The input is the source code of all methods in a
project. The output is the ranked list of methods with the detection
result/score and the interpretation (PDG sub-graph).
%Our solution is designed with
Let us explain our key ideas.

\subsubsection{{\bf Graph-based Vulnerability Detection Model} (Section~\ref{detect:sec})}
\label{vd-idea}

As seen in Section~\ref{motiv:sec}, a vulnerability is usually
exhibited as multiple statements are exploited, thus, it is natural to
capture the vulnerable code as a sub-graph in the PDG with the data
and control flows. To do so, we model the vulnerability
detection via the Graph Convolutional Network (GCN)~\cite{GCN16} as
follows. The PDG of a method $M$ is represented as a graph $G_N$ =
$(V,E)$ in which $V$ is a set of nodes representing the statements,
and $E$ is a set of edges representing the data/control dependencies.
A feature description $x_V$ is for every node $v$, which represents a
property of a node, e.g., variable name, etc. Features are summarized
in a $N \times D$ feature matrix $X_M$ ($N$: number of nodes and $D$
is the number of input features).
Let $f$ be a label function on the statements and methods $f: V
\rightarrow \{1,...,C\}$ that maps a node in $V$ and an entire method
to one of the $C$ classes. In {\tool}, $C$=2 for vulnerable
($\mathcal{V}$) and non-vulnerable ($\mathcal{NV}$).

%To adapt GCN for vulnerability detection, we use a node in the graph
%to represent a statement, and an edge for a~data~or control
%dependency among nodes. The features of a node~are used to model the
%properties of a program element or a~statement, e.g., the variables
%in a statement, the slice from a variable, etc. The label function
%$f$ maps each statement and the entire method to a set of two
%elements:~vulnerable or non-vulnerable.

%The GCN model is then trained on all the statements (nodes) in the
%training set and then used for prediction to approximate the label $f$
%on the graph $G_M$ of a given method $M$, i.e., to predict whether $M$
%is vulnerable or not. Specifically,

For training on (non-)vulnerable code in the training set, GCN
performs similar operations as CNN where it learns the features with a
small filter/window sliding over PDG sub-structure. Differing from
image data with CNN, the neighbors of a node in GCN are unordered and
variable in size. To predict if a method $M$ is vulnerable, its PDG
$G_M$ with the associated feature set $X_M$ = $\{x_j|v_j \in G_M\}$
are built. GCN learns a conditional distribution $P(Y|G_M,X_M)$, where
$Y$ is a random variable representing the labels $\{1,...,C\}$. That
distribution indicates the probability of the graph $G_M$ belonging to
each of the classes $\{1,...,C\}$, i.e., $M$ is vulnerable or not
(Section~\ref{detect:sec}).

\subsubsection{\bf Distinction between Vulnerable Statements and Surrounding Contexts} During training, for each vulnerable
statement $s$ in a method in the training dataset, we distinguish $s$
and the surrounding contextual statements for $s$. A context consists
of the statements with data and/or control dependencies with $s$.
%with a vector representation in the consideration with the vectors of
%surrounding statements that have data/control dependencies with $s$.
This is expected to help our model recognize better the vulnerable
code appearing in specific surrounding contexts, and have better
discriminating the vulnerable code from the benign one.
For example, the existing approaches feed the entire PDG of the
method in Figure~\ref{fig:pdg} into a model. {\tool} distinguishes and
learns the vector representation for the vulnerable statement at line
27 while considering as contexts the statements with data/control
dependencies with line 27: the data-dependency context (lines 31, 22,
13, 10, and 6), and the control-dependency context (lines 29, 25, 23,
and 13).

%in conjunction with the related statements via data/control
%dependencies with line 27, including lines 6, 22, 25, and 29.

%In addition, our model also considers the fixed statements and their
%program elements, e.g., the statement at line 28. This allows our
%model have better discrimination the vulnerable code from the benign
%one. Moreover, we also consider the code metrics~\cite{yi} that have
%shown to be crucial and discriminative in vulnerable code such as the
%number of nested code structures and iterations, etc.

\vspace{-2pt}
\subsubsection{{\bf Graph-based Interpretation Model for Vulnerability Detection} (Section~\ref{interpretation:sec})} 

%GNNEXplainer
After prediction, {\tool} performs fine-grained interpretation. It
uses both the PDG $G_M$ of the method $M$ and the GCN model as the
input to obtain the interpretation. To that end, we leverage the
interpretable ML technique {\em GNNExplainer}~\cite{GNNExplainer}. Its
goal is to take the GCN and a specific input graph $G_M$, and produce
the {\em crucial sub-graph structures and features} in $G_M$ that
affect the decision of the model. GNNExplainer's idea is that {\em if
  removing or altering a node/feature does affect the prediction
  outcome, the node/feature is considered as essential and thus must
  be included in the crucial set} (let us call it the {\em
  interpretation} set). GNNExplainer searches for a sub-graph
$\mathcal{G}_M$ in $G_M$ that minimizes the difference in the
prediction scores between using the whole graph $G_M$ and using the
minimal graph $\mathcal{G}_M$
(Section~\ref{interpretation:sec}). Because without that subgraph
$\mathcal{G}_M$ in the input PDG $G_M$, GCN model would not decide
$G_M$ as vulnerable, $\mathcal{G}_M$ is considered as crucial {\bf PDG
  sub-graph} consisting of {\bf crucial statements} and data/control
dependencies relevant to the detected vulnerability (if the outcome is
$\mathcal{V}$). If the outcome is non-vulnerability, $\mathcal{G}_M$
can be considered as the safe statements in PDG for the model to
decide the input method $M$ as benign code.

%% file: sections/appro.tex
\input{sections/approach_detection}

\input{sections/approach_interpretation}

%% file: sections/approach_detection.tex
\section{Graph-based Vulnerability Detection Model}
\label{detect:sec}

\subsection{Representation Learning}
\label{replearn:sec}

%Next, let us explain how we use FA-GCN for vulnerability
%detection.
Let us present how we build the vector representations for code
features.
%that we extract from source code.
%{\tool} first generates the PDG $G_M=(V,E)$ for method $M$. {\bf Each
%  node} represents a statement while {\bf each edge} is for a control
%or a data dependency.
For a {\em statement}, we extract the following types of {\bf
  features}:

%\begin{itemize}

%{\tool} uses a deep learning-based summarization approach \cite{} to
%summarize the sub-tree $Tree_{v}$ which represent the statement $v$
%in the whole AST for method $M$ into a feature representation vector
%$F_{v,1}$.

\begin{figure*}[t]
	\centering
	\includegraphics[width=5.8in]{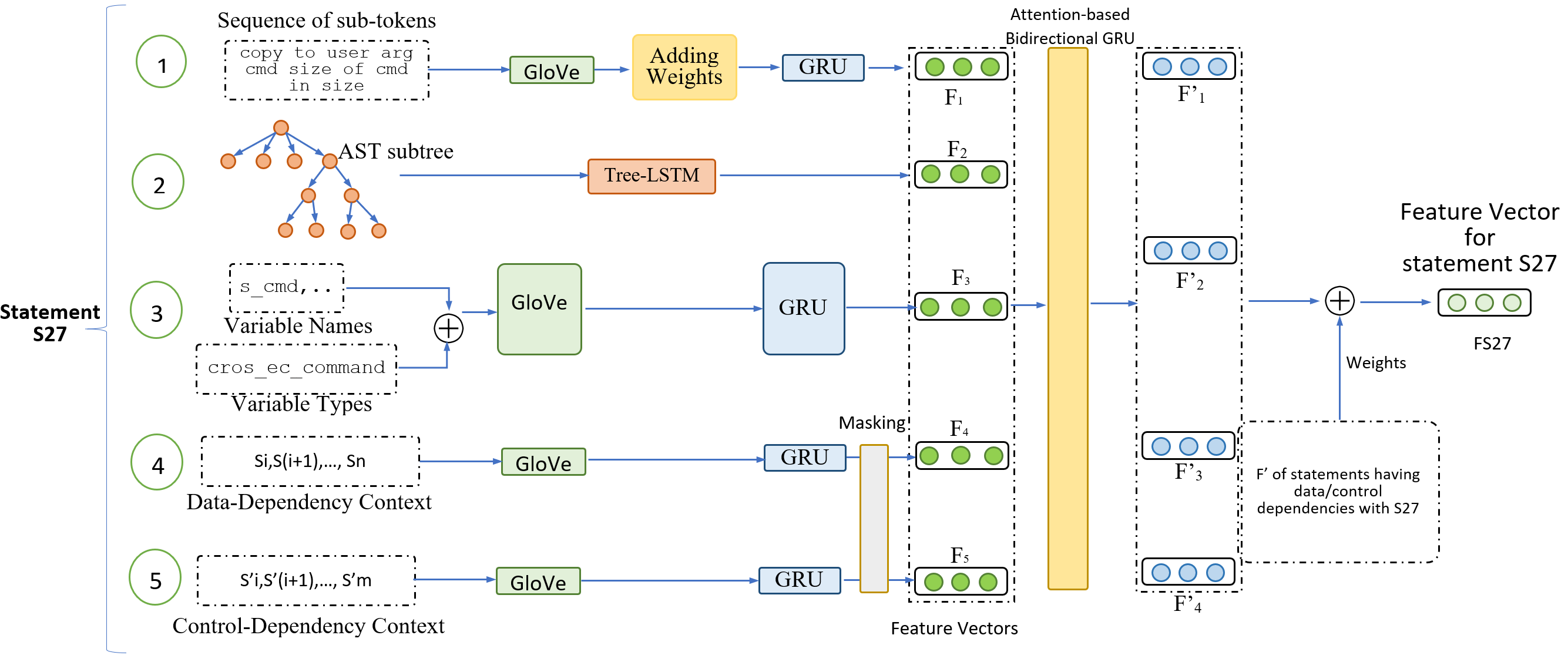}
        \vspace{-0.12in}
	\caption{Code Representation Learning for Statement S27 in Graph-based Vulnerable Code Detection}
%        \vspace{-0.05in}
	\label{fig:feature}	
\end{figure*}

\begin{figure*}[t]
	\centering
	\includegraphics[width=6in]{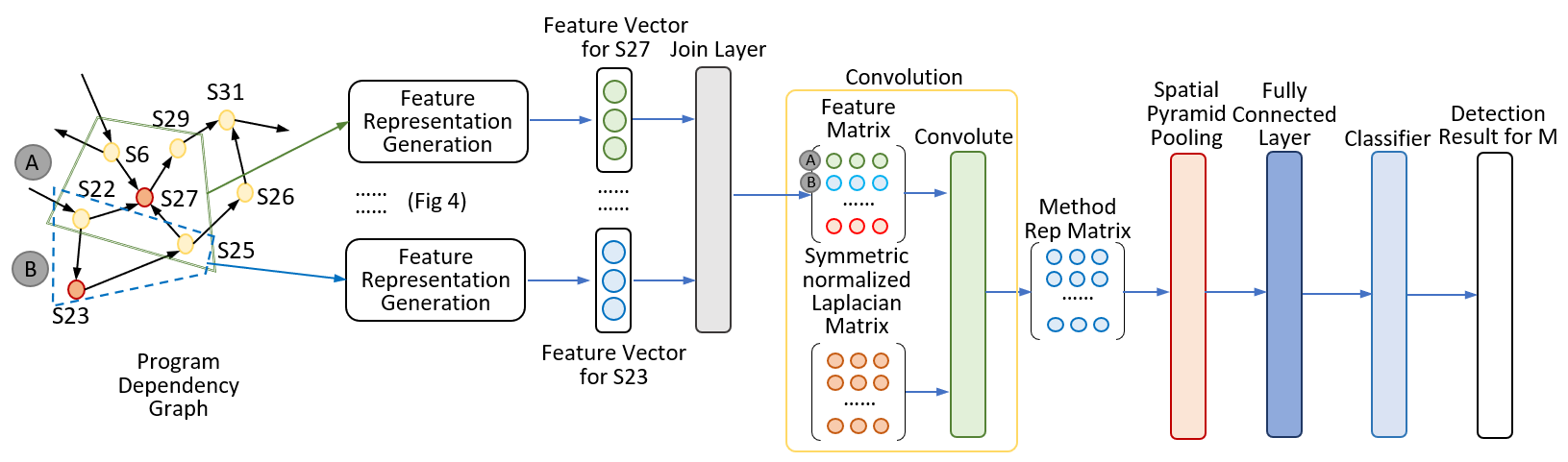}
  %      \vspace{-0.1in}
	\caption{Vulnerability Detection with FA-GCN}
	\label{fig:gcn}	
\end{figure*}

%\subsubsection{{\bf Sequence of sub-tokens.}}

\vspace{0.04in}
\noindent {\bf 1. Sequence of Sub-tokens of a Statement.}
At the lexical level, we capture the content of a statement in term of
the sequence of sub-tokens. We choose the sub-token granularity
because the sub-tokens are more likely to be repeated than the entire
lexical tokens in source code~\cite{icse20-methodname}.
%because they are basic elements in source code.
%To distinguish between the vulnerable code and the surrounding code,
%for a vulnerable statement, we also consider the fixed statement
%during training. For the non-vulnerable statements or deleted ones, we
%use the zero vector.
%
We tokenize each statement and keep only the variables, method and
class names. The names are broken into sub-tokens using CamelCase or
Hungarian convention. We remove the sub-tokens with one character to
avoid the influence of noises. For example, in
Figure~\ref{fig:feature}, the tokens of $S_{27}$ are collected and
broken down into the sequence: \code{copy}, \code{to}, \code{user},
\code{arg}, etc. Then, we use GloVe~\cite{glove2014}, to build the
vectors for tokens, together with Gate Recurrent Unit
(GRU)~\cite{chung2014empirical} to build the feature vector for the
sequence of sub-tokens for $S_{27}$.
GloVe is known to capture well semantic similarity among tokens. GRU
is chosen to summarize the sequence of vectors into one feature vector
for the next step.
%In the case of a vulnerable statement and its fixed ones, we combine
%them via multiplication to get the feature vector.

%is an effective method for measuring the linguistic or semantic
%similarity of the tokens. We need the GRU to summarize the sequence of
%vectors into one feature vector for the next step.

%{\tool} use GloVe here because we would like to use a word
%representation tool to transform the natural language tokens into
%vectors which could be used in GCN model. After apply GloVe, the
%sequence of tokens will be transformed into a sequence of vectors. In
%order to get only one vector for one feature, we use GRU here to
%summarize the sequence of vectors into one feature vector for the next
%step.

%{\tool} braking the statement $v$ into sequence and using a GRU model \cite{} to summarize the sequence as the second feature representation vector $F_{v,2}$ for statement $v$ by only taking variable, method, and class names and using the CamelCase and Hungarian convention to break each name into a sequence of sub-tokens and in order to avoid influence of basis, {\tool} removed one character length sub-tokens.

%\subsubsection{{\bf Code structure of the statement}}
%\label{ast:sec}

\vspace{0.05in}
\noindent {\bf 2. Code Structure of a Statement.}  We capture code
structure via the AST sub-tree.
%for each statement.
%Figure~\ref{fig:feature} shows the process of building feature
%representations for the statement $S_{27}$ at line 27.
In Figure~\ref{fig:feature}, the AST sub-tree for $S_{27}$ is
extracted and fed to Tree-LSTM~\cite{tai2015improved} to capture the
structure into a vector $F_2$.

%Tree-based LSTM model is known for capturing well tree structures.

%\subsubsection{{\bf Variables and Types}}
%\label{var:sec}

\vspace{0.05in}
\noindent {\bf 3. Variables and Types.}  For each node (i.e., a
statement), we collect the names of the variables and their static
types at their locations, break them into the sub-tokens. For example,
we collect the variable \code{s$\_$cmd} and its static type
\code{cross$\_$ec$\_$command}.
%Figure~\ref{fig:feature} displays the process (marked with
%\circled{3}) for the variable \code{s$\_$cmd}.  We collect two types
%of information for that variable: 1) its name, and 2) its static
%type.
%We concatenate all the sub-tokens in the names of variables and types
%into acsequence.
%
We use the same vector building techniques as for the sub-token
sequences as in feature 1, including GloVe and GRU, to apply on the
sequences of sub-tokens built from the variables' names (e.g.,
\code{s$\_$cmd}) and those from the variables' types (e.g,
\code{cross$\_$ec$\_$command}).

\vspace{0.05in}
\noindent {\bf 4. Surrounding Contexts.}
%\subsubsection{{\bf Surrounding Contexts (FIXME)}}
During training, for a statement $s$, we also encode the statements
surrounding $s$, which we refer to as {\em context}. Data- and
Control-dependency contexts contain the statements having such
dependencies with the current statement.
%We use two different criteria to define the context surrounding $s$,
%which can contain the statements with different types of relationships
%with $s$ (control dependencies, data dependencies).
For example, the data-dependency context for $S_{27}$ includes
%if data dependencies are considered, the statements that have data
%dependencies with $S_{27}$ are included in the context:
the statements at the lines 31, 22, 13, 10, and 6.  If the control
dependencies are considered, the statements with control dependencies
with $S_{27}$ at the lines 29, 25, 23, and 13 are included.
The vectors for the statements in the context are calculated via GloVe
and GRU as described earlier. The number of dependencies could be
different, then the lengths of the GRU model inputs could be
different. Therefore, we apply zero padding with a masking layer,
which allows the model to skip the zeros at the end of the sequence of
sub-tokens. Those zeros will not be included in the training.

%The masking layer we set to let the model avoid zero value which means
%when reach the end of the sequence of tokens, the rest zero padding
%part will not be include in the training and the model can jump these
%data to avoid the accuracy loss.

%$After GRUs, we apply a masking layer for
%the two contexts to play the role of the selection/combination.
%
%They are combined via ...  to make the feature vector for the context
%of the current statement under consideration.

\vspace{0.05in}
\noindent {\bf 5. Attention-based Bidirectional GRU.}
%\subsubsection{{\bf LSTM Model and Attention Layer}}
After having all vectors for the features $F_1$, $F_2$, ..., we use a
bi-directional GRU and an attention layer to learn the weight
vector $W_i$ for each feature $F_i$, based on the hidden states from
that model.  Then, we compute the weighted vector for each feature by
multiplying the original vector for the feature by the weight: $F'_i$
= $W_i$.$F_i$.

%After having all representation vectors for features, we regard them
%as a sequence of vectors and use a bi-directional LSTM model to learn
%the hidden statues $h_{S-27, j}$ for each feature $j$, where $j =
%{1,...,k+2}$. With an attention layer, {\tool} could get the weight
%vectors $W_{S-27, m}$ for each feature based on the hidden statues
%from bi-directional LSTM. In order to get the final feature
%representation vector $F_{S-27}$ for \textit{S-27}. {\tool} firstly
%calculate the weighted representation vectors $F'_{S-27, j}$ by:
%
%\begin{equation}\label{eq:8}
%F'_{S-27, j} = W_{S-27, j}F_{S-27, j}
%\end{equation}

%At the final step of feature representation building,

Finally, we need to consider the impacts from the {\em dependent statements to
  the current statement in the PDG}. The rationale is that those
neighboring statements in the PDG must have the influence on the
current statement if one of them is vulnerable. For example, the
neighboring statements for $S_{27}$ in the PDG include the statements
at lines 6, 22, 25, and 29. Thus, we combine and summarize them into
the final feature vector $F_{S27}$ for the statement $S_{27}$
%the statement at line 27
as follows:
\begin{equation}\label{eq:9}
F_{S27} = \sum_i{W_i{Concat(h(F'_i,j))}}
\end{equation}
$W_i$ is the trainable weight for combination; $Concat$ is the
concatenate layer to link all values into one vector; $h$ is the
hidden layer to summarize vector into a value; $i$ = S6, S22, S25,
S27, S29; $j$ is feature index. $F_{27}$ is used in the next step
with GCN model for detection.

\subsection{Vulnerability Detection with FA-GCN}
\label{model:sec}

Figure~\ref{fig:gcn} presents how we use Feature-Attention GCN model
(FA-GCN)~\cite{FAGCN} for detection. The rationale is that FA-GCN can
deal well with the graphs with sparse features (not all the statements
share the same properties), and potentially noisy features in a PDG.
%Figure~\ref{fig:gcn} explains how {\tool} detects vulnerable code using
%FA-GCN.
First, we parse the method $M$ into PDG. Similar to CNN using the
filter on an image, FA-GCN performs sliding a small window along all
the nodes (statements) of the PDG. For example, in
Figure~\ref{fig:gcn}, the window marked with \circled{A} for the node
$S27$ consists of itself and the neighboring statements/nodes $S6$,
$S22$, $S25$, and $S29$. Another window (marked with \circled{B}) is
for the node $S23$, including itself and the neighboring nodes:
$S22$ and $S25$. For each window, FA-GCN generates the feature
representation matrix for the statement at the center. For example,
for the window centered at $S27$, it generates the feature vector
$F_{S27}$ for $S27$, using the process explained in
Figure~\ref{fig:feature}. From the representation vectors for
all statements, FA-GCN uses a join layer to link all these vectors
into the Feature Matrix $\mathcal{F}_{m}$ for method $M$. A row in
$\mathcal{F}_m$ corresponds to a window in~PDG.

Next, FA-GCN performs the convolution operation by first calculating
the symmetric normalized Laplacian matrix~$\tilde{A}$~\cite{GCN16},
and then calculating the convolution to generate the representation
matrix $M_{m}$ for the method $m$. After that, we use the traditional
steps as in a CNN model: using a spatial pyramid pooling layer (to
normalize the method representation matrix into a uniform size, and
reduce its total size), and connecting its output to a fully connected
layer to transform the matrix into a vector $V_m$ to represent
$m$. With $V_m$, we perform classification by using two hidden layers
(controlling the length of vectors and output) and a softmax function
to produce a prediction score for $m$. We use those scores as {\em
  vulnerability scores to rank the methods} in a project. The decision
for $m$ as $\mathcal{V}$ or $\mathcal{NV}$ is done via a trainable
threshold on the prediction score~\cite{li2018vuldeepecker,li2019improving}.

%% file: sections/approach_interpretation.tex
\section{Graph-based Interpretation Model}
\label{interpretation:sec}

\begin{figure}[t]
	\centering
	\includegraphics[width=3in]{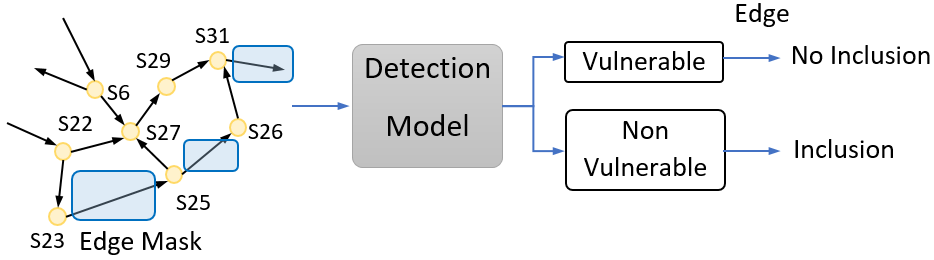}
        \vspace{-0.12in}
	\caption{Masking to Derive Interpretation Sub-Graphs}
        \vspace{-0.06in}
	\label{fig:GNNEX}	
\end{figure}

Let us explain how we use GNNExplainer~\cite{GNNExplainer} to build our graph-based
interpretation. The input includes the trained FA-GCN model, the PDG
($G_M$) of the method $M$, and the detection result $\mathcal{V}$ or
$\mathcal{NV}$, and prediction score. Figure~\ref{fig:GNNEX}
illustrates our process for the case of $\mathcal{V}$ (Vulnerable)
(the case of $\mathcal{NV}$ is done similarly).

%Let us explain how we use GNNExplainer to build our graph-based
%interpretation model. The input includes 1) the trained FA-GCN
%detection model (Section~\ref{detect:sec}), 2) the given PDG ($G_M$)
%of the method $M$, and 3) the dxsdetection result $\mathcal{V}$ or
%$\mathcal{NV}$. Fig.~\ref{fig:GNNEX} illustrates our process when the
%result is $\mathcal{V}$ (Vulnerable) (the case of $\mathcal{NV}$ is
%done similarly).

%The interpretations include 1) a crucial sub-graph $\mathcal{G}_M$,
%corresponding to the PDG sub-graph consisting of statements relevant
%to the vulnerability, and 2) a subset of crucial features
%$\mathcal{X}_M$, corresponding to the set of variables involving to
%the vulnerability.

To derive the interpretations, the key goal is to find a sub-graph
$\mathcal{G}_M$ in the PDG $G_M$ of the method $M$ that minimizes the
difference in the prediction scores between using the entire graph
$G_M$ and using the minimal graph $\mathcal{G}_M$. To do so, we use
GNNExplainer with the {\em masking technique}~\cite{GNNExplainer},
which treats the searching for the minimal graph $\mathcal{G}_M$ as a
learning problem of the {\em edge-mask} set $EM$ of the edges.
% two sets: the {\em edge-mask} set $EM$ of edges and the {\em
%feature-mask} set $FM$ of features.
The idea is that learning $EM$
%these two sets $EM$ and $FM$
helps {\tool} derive the interpretation sub-graph $\mathcal{G}_M$
%and the set of crucial feature $\mathcal{X}_M$
by masking-out the edges in $EM$
%and the feature in~$FM$ from $G_M$ and $X_M$:
from $G_M$ (``masked-out'' is denoted by $\bigodot$):
\begin{equation}\label{eq:11}
\mathcal{G}_M = G_M \bigodot EM
\end{equation}
Figure~\ref{fig:GNNEX} illustrates GNNExplainer's principle. As an
edge-mask set is applied, GNNEXplainer checks if the FA-GCN model
produces the same result~(in this case the result is
$\mathcal{V}$). If yes, the edge in the edge-mask is not important and
is not included in $\mathcal{G}_M$. Otherwise, the edge is important
and included in $\mathcal{G}_M$.
%Similar treatment is for {\em feature-mask} $FM$. In
%Figure~\ref{fig:GNNEX}, $FM$ masks the second and fourth features for
%each statement (note: each statement has the same number of
%features).
Because the numbers of possible sub-graphs and the edge-mask sets
%and feature-mask
are untractable, GNNExplainer uses a learning approach for the
edge-mask $EM$.

%and $FM$.

%\begin{equation}\label{eq:12}
%\mathcal{X}_M = X_M \bigodot FM
%\end{equation}

%For example, as you can see in figure \ref{fig:GNNEX}, when the
%GNNExplainer get the whole PGD $G_m$, it starts to use the
%\textit{edge-mask} $EM$ and \textit{feature-mask} $FM$ to mask some
%edges and features.  As you can see in the second row of the figure,
%after masked some edges from $G_m$ to generate a graph $G'_m$,
%GNNExplainer use the detection model to check the new graph to see
%how the detection result changes. If the result has been influenced a
%lot, it means that the masked edges are important for the detection
%result. If not, it means that the masked edges are not important for
%the detection result. It is similar for the \textit{feature-mask}
%$FM$.  Because in the last step, we could get $F_{v}$ for each
%statement $v$ and $F_{v}$ have the same dimension, $FM$ masks the
%features in the same position in $F_{v}$ for each statement $v$ and
%then use detection model to evaluate the importance of the masked
%features just as the edges. As you can see in figure \ref{fig:GNNEX},
%$FM$ masked the second and the 4th feature for each statement in this
%example.

%\begin{align}\label{maineq}
%\nonumber
%\max_{\mathcal{G}_M} MI(Y,(\mathcal{G}_M,\mathcal{X}_M)) = H(Y) - H(Y|G=\mathcal{G}_M,X=\mathcal{X}_M)
%\end{align}

%to minimizing conditional entropy
%$H(Y|G=\mathcal{G}_M,X=\mathcal{X}_M)$

%\begin{equation}
%  \label{eq2}
%-\EX_{Y|\mathcal{G}_M,
%  \mathcal{X}_M} [log P_{FA-GCN} (Y|G=\mathcal{G}_M,X=\mathcal{X}_M)]
%  \end{equation}

%\begin{equation}
%  \label{eq3}
%  \min_{\mathcal{G}} \EX_{\mathcal{G}_M \sim \mathcal{G}} H(Y|G=\mathcal{G}_M,X=\mathcal{X}_M)
%\end{equation}
%\begin{equation}
%  \label{eq4}
%  \min_{\mathcal{G}} H(Y| G=\EX_{\mathcal{G}}[\mathcal{G}_M], X=\mathcal{X}_M)
%\end{equation}
%\nonumber

Let us formally explain how GNNEXplainer~\cite{GNNExplainer} works. It
formulates the problem by maximizing the mutual information (MI)
between the minimal graph $\mathcal{G}_M$ and the input~$G_M$:
\begin{equation}\label{maineq}
\max_{\mathcal{G}_M} MI(Y,\mathcal{G}_M) = H(Y) - H(Y|G=\mathcal{G}_M)
\end{equation}
$Y$ is the outcome decision by the FA-GCN model. Thus, the entropy term
$H(Y)$ is constant for the trained FA-GCN model. Maximizing the $MI$
value for all $\mathcal{G}_M$ is equivalent to minimizing conditional
entropy $H(Y|G=\mathcal{G}_M)$, which by definition of
conditional entropy can be expressed~as
\begin{equation}
  \label{eq2}
-\EX_{Y|\mathcal{G}_M}
  [log P_{FA-GCN} (Y|G=\mathcal{G}_M]
  \end{equation}
The meaning of this conditional entropy formula is a measure of how
much uncertainty remains about the outcome $Y$ when we know
$G=\mathcal{G}_M$. GNNEXplainer also limits the
size of $\mathcal{G}_M$ by $K_M$, i.e., taking $K_M$ edges that give
the highest mutual information with the prediction outcome $Y$.
Direct optimization of the formula~\ref{eq2} is not tractable, thus,
GNNExplainer treats $\mathcal{G}_M$ as a random graph variable
$\mathcal{G}$. The objective in Equation~\ref{eq2} becomes:
\begin{equation}
  \label{eq3}
  \min_{\mathcal{G}} \EX_{\mathcal{G}_M \sim \mathcal{G}} H(Y|G=\mathcal{G}_M)
\end{equation}
\begin{equation}
  \label{eq4}
  \min_{\mathcal{G}} H(Y| G=\EX_{\mathcal{G}}[\mathcal{G}_M])
\end{equation}
From Equation~\ref{eq3}, we obtain Equation~\ref{eq4} with Jensen's
inequality.  The conditional entropy in Equation~\ref{eq4} can be
optimized by replacing $\EX_{\mathcal{G}}[\mathcal{G}_M]$ to be
optimized by masking with $EM$ on the input graph $G_M$.
Now, we can reduce the problem to learning the mask $EM$.
%Similar treatment is applied to $FM$.
Details on training can be found in~\cite{GNNExplainer}. The resulting
sub-graph $\mathcal{G}_M$ is directly used as an interpretation. 
We can similarly produce the interpretations for the cases of non-vulnerability result.

%% file: sections/exper.tex
\section{Empirical Evaluation}
\label{sec:eval}

\input{sections/RQs}
\input{sections/exper_methodology}

\input{sections/exper_results}

%% file: sections/RQs.tex
\subsection{Research Questions}

To evaluate {\tool}, we seek to answer the following questions:

\noindent\textbf{RQ1. Comparison on the Method-Level Vulnerability
  Detection (VD).}  How well does {\tool} perform in comparison with
the state-of-the-art method-level Deep Learning VD approaches?

\noindent\textbf{RQ2. Comparison with other Interpretation Models for
  Fine-grained VD Interpretation.}  How well does {\tool} perform in
comparison with the state-of-the-art interpretation models for
fine-grained VD interpretation to point out vulnerable statements?

\noindent\textbf{RQ3. Vulnerable Code Patterns and Fixing Patterns.}
Is {\tool} useful in detecting vulnerable code patterns and fixes?

\noindent\textbf{RQ4. Sensitivity Analysis for Internal Features.} How
do internal features affect the overall performance of {\tool}?

%\noindent\textbf{RQ4. Sensitivity Analysis for Internal Parameters.}
%How do key internal parameters affect the performance of {\tool}?

\noindent\textbf{RQ5. Sensitivity Analysis on Training Data.}  How do
different data splitting schemes affect {\tool}'s performance?

\noindent\textbf{RQ6. Time Complexity.} What
is time complexity of {\tool}?

%\noindent\textbf{RQ1. Accuracy in Interpretations for VD.}  How well
%does {\tool} provide interpretation for vulnerability detection?

%\noindent\textbf{RQ2. Accuracy on the CWE retrieval.}  How well does
%{\tool} perform on retrieving relevant CWE for a given code?

%\noindent\textbf{RQ3. Comparison on the DL-based Vulnerability
%  Detection.}  How well does {\tool}
%perform in comparison with the state-of-the-art DL-based
%VD approaches?

%\noindent\textbf{RQ3. Performance of the Enhanced {\tool}.} 
%Can we improve the overall performance of {\tool} in APR?

%\noindent\textbf{RQ4. Sensitivity Analysis of {\tool}.} 
%How do various factors affect the overall performance of {\tool}?

%\noindent\textbf{RQ4. Case Study.} 
%How does {\tool} process and perform on a real vulnerability?

%% file: sections/exper_methodology.tex
%\subsection{Experimental Methodology}

\input{sections/dataset.tex}

\subsection{Experimental Methodology}
%We use the following analysis approaches to answer each RQ.

%\vspace{3pt}
\noindent\textbf{RQ1. Comparison on Method-Level DL-based VD Approaches.}

\textit{\underline{Baselines.}} We compare
{\tool} with the state-of-the-art DL-based vulnerability detection
approaches: 1) \textbf{VulDeePecker}~\cite{li2018vuldeepecker}: a
DL-based approach using Bidirectional LSTM on the statements and their
data/control dependencies. 2) \textbf{Devign}~\cite{zhou2019devign}:
an DL-based approach that uses GGCN model with Gated Graph Recurrent
Layers on the AST, CFG, DFG, and code sequences for graph
classification. 3) \textbf{SySeVR}~\cite{li2018sysevr}: in addition to
statements and program dependencies, this approach also uses program
slicing and leverages several DL models (LR, MLP, DBN, CNN, LSTM,
etc.). 4) \textbf{Russell} {\em et al}.~\cite{russell2018automated}: This DL
approach encodes source code as matrices of code tokens and leverages
convolution model with random forest (RF) via ensemble classifier.
5) \textbf{Reveal}~\cite{chakraborty2020deep}: This approach uses
GGNN, MLP, and with Triplet Loss on graph representations of
source code.

\textit{\underline{Procedure.}}
A dataset contains a number of vulnerable and non-vulnerable methods.
We first randomly split all of its vulnerable methods into 80\%, 10\%,
and 10\% to be used for training, tuning, and testing, respectively.
For training, we add to that 80\% part the same number of
non-vulnerable methods as the vulnerable ones to obtain the balanced
training data. For tuning and testing, we also add the non-vulnerable
methods but we use the real ratio between vulnerable and
non-vulnerable methods in the original dataset to build tuning/testing
data. We use AutoML~\cite{NNI} on all models to automatically tune hyper-parameters
on the tuning dataset.

We also performed the evaluation across the datasets. We first trained
our model on the combination of two datasets \code{Reveal}
and \code{FFMPeg+Qemu}, which has a balanced number of vulnerable
methods and non-vulnerable ones. We then tested the model
on \code{Fan} dataset, which has a more realistic ratio of vulnerable
and non-vulnerable methods. To ensure the model suitable for
cross-data evaluation, we also used 20\% of \code{Fan} dataset for
tuning the parameters and performed prediction on the remaining 80\%.

%We also perform cross-validation to test if our model has been trained
%on one dataset, how it performs on the other dataset. To do so, we
%first train the FA-GCN model on the combination of REVEAL dataset and
%FFMPeg+Qemu dataset which has a balanced number of vulnerable methods
%and non-vulnerable ones, and doing the prediction on the Fan {\em et
%al.}~\cite{fan2020msr} dataset which is in the real ratio between
%vulnerable and non-vulnerable methods in this dataset. To ensure the
%model is suitable for the cross-data setting, we picked 20\% of the
%Fan {\em et al.}~\cite{fan2020msr} dataset for tuning the parameters
%and doing prediction on the rest 80\%.

%To analysis the results between each baseline and our approach, we do the overlapping analysis in this RQ also. To be more in detail, after we run all approaches to do the prediction, we pick the top 100 methods that the model predicted as vulnerable for each approach and do the overlapping analysis between each baseline and our approach to analyze the performance difference.

\textit{\underline{Evaluation Metrics:}} We use the following evaluation metrics.

{\bf {\em Mean Average Precision}} $MAP$ $=\frac{\sum_{q=1}^{Q}AvgP(q)}{Q}$,
with {\em Average Precision} $AvgP$ $=\sum_{k=1}^{n}P(k)rel(k)$,
where $n$ is the total number of results $k$ is the current rank in
the list, $rel(k)$ is an indicator function equaling to 1 if the item
at rank $k$ is actually vulnerable, and to zero otherwise. $Q$ is the
total number of classification types. It is $1$ because we only have
two types including vulnerable and non-vulnerable classes, however, we
rank all the methods based on their scores (1 indicates vulnerable,
and 0 otherwise).

%we only care about the vulnerable results.

{\bf {\em Normalized DCG}} at $k$: {\bf $nDCG_k$}
$=\frac{DCG_k}{IDCG_k}$, with {\em Discounted Cumulative Gain} at rank
$k$, $DCG_k$ $=\sum_{i=1}^{k}\frac{r_i}{log_2(i+1)}$; and {\em Ideal
DCG} at $k$ $IDCG_k$
$=\sum_{i=1}^{|R_k|}\frac{2^{r_i}-1}{log_2(i+1)}$; where $r_i$ is the
score of the result at position $i$, and $R_k$ the rank of the actual
vulnerable methods (ordered by their scores) in the resulting list up
to the position $k$.

{\bf First Ranking} ($FR$) is the rank of the first correctly
predicted vulnerable method. {\bf Average ranking} ($AR$) is the
average rank of the correctly predicted vulnerable
methods in the top-ranked list.

{\bf Accuracy under curve} (AUC) is defined as $AUC = P(d(m_1)>d(m_2))$ in
which $P$ is the probability, $d$ is the detection model (can be
regarded as a binary classifier), $m_1$ is a randomly chosen positive
instance, and $m_2$ is a randomly chosen negative instance.

{\bf Precision} (P) is the fraction of relevant instances among the retrieved ones. It is calculated as $Precision = \frac{TP}{TP+FP}$ while $TP$ is the number of true positives and the $FP$ is the number of false positives.

{\bf Recall} (R) is the fraction of relevant instances that were retrieved. It is calculated as $Recall = \frac{TP}{TP+FN}$ while $TP$ is the number of true positives and the $FN$ is the number of false negatives.
 
{\bf F score} (F) is the harmonic mean of precision and recall. It is calculated as $F score = 2\frac{Precision*Recall}{Precision+Recall}$.

\vspace{0.03in}
\noindent\textbf{RQ2. Comparison with other Interpretation Models for Fine-grained Interpretation.}

\textit{\underline{Baselines.}} We compare {\tool} with the following interpretation models.
1) \textbf{ATT}~\cite{GNNExplainer}: This approach is a graph
attention network that uses the attention mechanism to evaluate the
weights (importance levels) of the edges in the input graph.
2) \textbf{GRAD}~\cite{GNNExplainer}: This approach is a
gradient-based method that computes the gradient of the GNN's loss
function {\em w.r.t.} the adjacency matrix.

\textit{ \underline{Procedure}.}
Our goal here is to evaluate how well {\tool} produces the
fine-grained interpretations pointing to vulnerable statements.  Thus,
to train/test the interpretation model,
%(Section~\ref{interpretation:sec}),
we need to use the \code{Fan} dataset because it contains the
vulnerable statements and respective fixes. The other two datasets
contain only the vulnerabilities at the method level and no fixes.
Therefore, in this RQ2, for the vulnerability prediction part, we used
the GCN-FA model that was trained on \code{Reveal}
and \code{FFMPeg+Qemu} and predicted on the \code{Fan} dataset. For
the methods that are vulnerable, but predicted as non-vulnerable, we
considered those cases as incorrect because the resulting
interpretations do not make sense for incorrect detection. 
For the
methods that are actually non-vulnerable (regardless of the
predictions), we could not use them because the non-vulnerable methods do
not have the fixed statements as the ground truth for
interpretations. Thus, we use the set of methods that are vulnerable
and correctly detected as vulnerable for the evaluation of the
interpretation model. Let us use $D$ to denote this set.

For the interpretation, we randomly split $D$ into 80\%, 10\%, and
10\% for training, tuning, and testing.
For training, we used {\em the fixed statements} as the labels for
interpretation because those fixed ones were the vulnerable ones. For
testing, we compared the relevant statements from the interpretation
model against the actual fixed statements. Each method in the testing
set and the trained GCN-FA model are the input of the interpretation
model in this RQ2.

{\it \underline{Evaluation Metrics}}. Given an interpretation
sub-graph $\mathcal{G}_M$
%containing $N$ nodes and $E$ edges
generated from the graph-based interpretation model, we evaluate the
accuracy of the interpretation for a model as follows. For a method,
if $\mathcal{G}_M$
%with $N$ nodes
has an overlap with any statement in the code changes that fix the
vulnerability, $\mathcal{G}_M$ is considered as a correct
interpretation, i.e., relevant to the VD. We then
calculate \code{Accuracy} as the ratio between the number of correct
interpretations over the total number of interpretations.
%We use \code{Accuracy} =
%$\frac{|Relevant\:\mathcal{G}_M|}{|Total\:\mathcal{G}_M|}$.
Because code changes could include addition, deletion, and
modification, we further define such overlap as follows.

%We calculate \code{Accuracy} of the interpretation results for the
%vulnerability cases (i.e., code is $\mathcal{V}$, and {\tool} predicts
%$\mathcal{V}$) as follows.
If one of the statements $S$ in the vulnerable version was {\em
deleted} or {\em modified} for fixing, and if $\mathcal{G}_M$ $\ni$
$S$, then we consider the interpretation sub-graph $\mathcal{G}_M$ is
correct, otherwise incorrect. If one of the statements $S'$ was {\em
added} to the vulnerable version for fixing, we check on the fixed
version whether $\mathcal{G}_M$ contains any statement with data or
control dependencies with $S'$, we consider it as correct, otherwise,
incorrect. For example, in Fig.~\ref{fig:pdg}, $\mathcal{G}_M$
contains the statement S23 with data and control dependencies with one
of the added lines from 17--21. Thus, $\mathcal{G}_M$ is correct.
%The reason is that $\mathcal{G}_M$ is the PDG sub-graph of the
%vulnerable version and $S'$ appears only in the fixed~version.
The rationale is that if the interpretation sub-graph $\mathcal{G}_M$
contains some statement relevant to the added statement to fix the
vulnerability, that interpretation is useful in pointing out the code
relevant to the vulnerability.

%NEW
%To compare the overall performance of the interpretation statements,

We also use Mean First Ranking (MFR), i.e., the mean of the rankings
for the first statement that needs to be fixed in the interpretation
statements, and Mean Average Ranking (MAR), i.e., the mean of the
rankings for all statements to be fixed in the interpretation
statements. If a statement to be fixed has not been selected as
interpretation, we do not consider it when calculating MFR/MAR.

%\underline{Interpretation Accuracy for Non-Vulnerability Cases:}
%We use the corresponding fixed versions of the vulnerable code as the
%test set for these cases. {\bf Yi: how do you have the fixes if they
%are non-vulnerable cases?}

%We calculate the \code{Accuracy} of the interpretation results for the
%non-vulnerability cases (i.e., code is $\mathcal{NV}$, and {\tool}
%predicts $\mathcal{NV}$) as follows.  If one of the statements $S'$ in
%the fixed version was {\em added} to the vulnerable version or {\em
%modified} for fixing, and if $\mathcal{G}_M$ $\ni$ $S'$, then we
%consider the interpretation sub-graph $\mathcal{G}_M$ is
%relevant. That is, $\mathcal{G}_M$ contains the added statements for
%fixing.  Otherwise, it is irrelevant. If one of the statements $S$ was
%{\em deleted} from the vulnerable version for fixing, we check in the
%vulnerable version whether $\mathcal{G}_M$ contains any statement with
%data/control dependencies with $S$, we consider it as~relevant,
%otherwise, irrelevant. The reason is that $\mathcal{G}_M$ is the PDG
%sub-graph of the fixed version and $S$ appears only in the vulnerable
%version.

\vspace{0.02in}
\noindent\textbf{RQ3. Vulnerable Code Patterns and Fixing Patterns.}

\textit{ \underline{Procedure.}} We use a mining algorithm
on the set of interpretation sub-graphs to mine patterns of vulnerable
code. We also mine fixing patterns for those vulnerabilities. See
details in Section~\ref{pattern:sec}.

{\it \underline{Evaluation Metrics}}. We counted the identified patterns.

\vspace{0.03in}
\noindent\textbf{RQ4. Sensitivity Analysis for Features.}

\textit{ \underline{Procedure.}}
We first built a base model with only the feature that represents the
code as the sequence of tokens. We then built other variants of our
model by gradually adding one more feature in Section~\ref{replearn:sec} to
the base model including the sequence of sub-tokens, AST subtree,
variable names, data dependencies, and control dependencies. We
measured accuracy for each variant. We used the \code{Fan} dataset and
the same experiment setting as in RQ1.

%In this RQ, we will evaluate the impact of each node feature in our model. We use the feature that is the natural sequence of the token as the base model. For each experiment, we add one more feature to the base model until all features have been added. We compare the improvements by adding each feature to evaluate the impact for them. We use the Fan {\em et al.}~\cite{fan2020msr} to do the evaluation in this RQ.

{\it \underline{Evaluation Metrics}}. We use the same metrics as in RQ1.

\vspace{0.03in}
\noindent\textbf{RQ5. Sensitivity Analysis for Training Data.}
We used different ratios in data splitting for training, tuning, and
testing: (80\%, 10\%, 10\%), (70\%, 15\%, 15\%), (60\%, 20\%, 20\%),
and (50\%, 25\%, 25\%). We used the same \code{Fan} dataset and
setting as in RQ1.

%In this RQ, we do the experiments to evaluate the influence of the different data splitting. We aim to try the data
%splittings as 80\%/10\%/10\% (Training/Tuning/Testing),
%70\%/15\%/15\%, 60\%/20\%/20\%, 50\%/25\%/25\%. We do this evaluation
%on Fan {\em et al.}~\cite{fan2020msr} dataset.

%The second one is the
%cross dataset validation which we train our model on one dataset and
%test on the other one to evaluate the cross method performance of our
%model. In this RQ, we train our model on the Reveal dataset and test
%the model on the Fan {\em et al.}~\cite{fan2020msr} dataset.

{\it \underline{Evaluation Metrics}}.  We use the same metrics as in
RQ1.

%\vspace{0.02in}
\noindent\textbf{RQ6. Time Complexity Analysis.} We measure
the actual training and predicting time.

%% file: sections/dataset.tex
%\vspace{-0.12in}
\subsection{Datasets}
\label{dataset}

\begin{table}[t]
	\centering
	\caption{Three Datasets}
	\vspace{-0.06in}
	\small
	\renewcommand{\arraystretch}{1}
	\begin{tabular}{|r|r|r|r|}
		\hline
		Dataset  & Fan & Reveal & Devign \\
		\hline
		Vulnerabilities   & 10,547 & 1,664 & 10,067 \\
		Non-vulnerabilities           & 168752 & 16505 & 12,294\\
		Ratio (Vul:Non-vul)   & 1:16 & 1:9.9 & 1:1.2 \\
		%...     &  &  &  \\
		\hline
	\end{tabular}
	\label{datasets}
%	\vspace{-0.06in}
\end{table}

We have conducted our study on three vulnerability datasets including Fan
{\em et al.}'s~\cite{fan2020msr}, Reveal~\cite{chakraborty2020deep}
and FFMPeg+Qemu~\cite{zhou2019devign} (Table~\ref{datasets}).
%
%3754 code vulnerabilities spanning 91 different vulnerability
%types. All these code vulnerabilities are extracted from 348 Git
%projects.
%
Fan {\em et al.}~\cite{fan2020msr} dataset covers the CWEs from
2002 to 2019 with 21 features for each vulnerability.
%There are 3754 code vulnerabilities spanning 91 different vulnerability types, which were extracted from 348 Git projects. 
At the method level, the dataset contains +10K vulnerable methods and
fixed code. The Reveal dataset~\cite{chakraborty2020deep} contains
+18K methods with 9.16\% of the vulnerable ones. The FFMPeg+Qemu dataset
has been used in Devign study~\cite{zhou2019devign} with +22K data,
and 45.0\% of the entries are vulnerable.

%\\\\\\le~\ref{datasets} shows the statistics of three datasets.

%The vulnerable methods are used as the test set to evaluate {\tool}'s accuracy in detecting vulnerable code, while the fixed versions of those methods are used as the test set to evaluate its accuracy in recognizing the healthy code.
%
%Because our approach is working on the method level, we break down the
%dataset into method level by splitting each changed method in the
%vulnerability fixing commits as one vulnerable method to deal with. A
%special situation we considered is that we marked all changed testing
%method as non-vulnerable because they are often changed to match the
%vulnerability fixing, but they often are not about the
%vulnerability. With the process like this, we can get 10,000+
%vulnerable methods in total for our empirical evaluation.

%% file: sections/exper_results.tex
\section{Experimental Results}

\input{sections/rq1-results}

\input{sections/rq2-results}

\input{sections/rq4-results}

\input{sections/rq3-results}

\input{sections/rq5-results}

\input{sections/rq6-results}

%% file: sections/rq1-results.tex
\subsection{{\bf RQ1. Comparison on Method-Level VD}}

\begin{table}[t]
	\caption{RQ1. Top-10 Vulnerability Detection Ranked Results on
	FFMPeg+Qemu Dataset. 0: incorrect, 1:
	correct}
        \vspace{-0.1in}
        \begin{center}
          \footnotesize
          \tabcolsep 4pt
	 \renewcommand{\arraystretch}{1} \begin{tabular}{p{1.5cm}<{\centering}|p{0.3cm}<{\centering}|p{0.3cm}<{\centering}|p{0.3cm}<{\centering}|p{0.3cm}<{\centering}|p{0.3cm}<{\centering}|p{0.3cm}<{\centering}|p{0.3cm}<{\centering}|p{0.3cm}<{\centering}|p{0.3cm}<{\centering}|p{0.3cm}<{\centering}|p{0.5cm}<{\centering}}
		
	\hline
Top-10 result &1 & 2 &3&4&5&6&7&8&9&10 & Total\\
\hline
VulDeePecker         &0&0&0&0&0&0&{\bf 1} &0&1&1 & 3 \\
SySeVR &0&0&0&0&0&{\bf 1}&1&1&0&1 & 4 \\
Russell {\em et al.} &0&0&0&0&{\bf 1}&0&1&0&1&1 & 4\\
Devign &0&0&0&0&{\bf 1}&0&1&1&1&0 & 4\\
Reveal  &0&0&0&{\bf 1}&0&1&0&1&1&1 & 5\\
%\hline
{\tool} &{\bf 1}&0&1&1&1&0&1&1&0&0 & 6\\
\hline
		\end{tabular}
	\label{RQ1-data-3-top10}
        \end{center}
        \vspace{-0.12in}
\end{table}

\begin{table}[t]
  \caption{RQ1. Method-Level VD on FFMPeg+Qemu Dataset}
  \vspace{-0.1in}
	\begin{center}
		\footnotesize
		\renewcommand{\arraystretch}{1}
		\begin{tabular}{p{1.1cm}<{\centering}|p{1cm}<{\centering}|p{0.7cm}<{\centering}|p{0.7cm}<{\centering}|p{0.7cm}<{\centering}|p{0.7cm}<{\centering}|p{1cm}<{\centering}}
			\hline
			&  VulDee- -Pecker &  SySeVR  &  Russell {\em et al.}  & Devign     & Reveal        & {\tool}    \\
			\hline
			nDCG@1&       0 &0&0&0&0&1 \\
			nDCG@3&       0 &0&0&0&0&0.63 \\
			nDCG@5&       0 &0&0.43&0.45&0.5& 0.65\\
			nDCG@10&      0.37  &0.44&0.45&0.46&0.5& 0.68\\
			nDCG@15&        0.45&0.48&0.49&0.52&0.55& 0.75\\
			nDCG@20&        0.48&0.51&0.54&0.56&0.6& 0.82\\
			\hline
			MAP@1&    0    &0&0&0&0&1 \\
			MAP@3&    0    &0&0&0&0& 0.83\\
			MAP@5&    0    &0&0.20&0.20&0.25& 0.80\\
			MAP@10&    0.22    &0.31&0.30&0.32&0.38& 0.78\\
			MAP@15&      0.29  &0.33&0.34&0.37&0.41& 0.72\\
			MAP@20&       0.32 &0.35&0.37&0.42&0.45& 0.69\\
			\hline
			FR@1&        n/a&n/a&n/a&n/a&n/a& 1\\
			FR@3&        n/a&n/a&n/a&n/a&n/a& 1\\
			FR@5&        7&6&5&5&4& 1\\
			FR@10&        7&6&5&5&4&1 \\
			FR@15&        7&6&5&5&4&1 \\
			FR@20&        7&6&5&5&4&1 \\
			\hline
			AR@1&       n/a &n/a&n/a&n/a&n/a& 1\\
			AR@3&        n/a&n/a&n/a&n/a&n/a& 2\\
			AR@5&        n/a&n/a&5&5&4& 3.3\\
			AR@10&       8.7 &7.8&7.8&7.4&7.4& 4.7\\
			AR@15&        11.2&10&9.5&10&9.1& 7.6\\
			AR@20&        13.3&12.1&12.6&12.1&12.4& 10.3\\
			\hline
			AUC&        0.68&0.72&0.79&0.77&0.79& 0.84\\
			\hline
		\end{tabular}
		\label{RQ1-data-3}
	\end{center}
        \vspace{-0.12in}
\end{table}

%\begin{figure}[t]
%	\centering
%	\subfloat[MAP Scores]{{\includegraphics[width=0.235\textwidth]{graphs/MAP_2.png}}}
%	\subfloat[nDCG Scores]{{\includegraphics[width=0.235\textwidth]{graphs/nDCG_2.png}}}
	%	\begin{minipage}[t]{0.4\textwidth}
	%		\centering
	%		\includegraphics[width=0.4\textwidth]{graphs/MAP_1.png}
	%		\caption{MAP Scores}
	%		\label{RQ1-map-1}
	%	\end{minipage}
	%	\begin{minipage}{0.4\textwidth}
	%	\centering
	%	\includegraphics[width=0.4\textwidth]{graphs/nDCG_1.png}
%	\caption{Scores Changes from Top 1 to Top 100 on Reveal Dataset}
%	\label{RQ1-score-2}
	%\end{minipage}
%\end{figure}

%\begin{figure}[ht]
%	\centering
%	\subfloat[MAP Scores]{{\includegraphics[width=0.235\textwidth]{graphs/MAP_3.png}}}
%	\subfloat[nDCG Scores]{{\includegraphics[width=0.235\textwidth]{graphs/nDCG_3.png}}}
	%	\begin{minipage}[t]{0.4\textwidth}
	%		\centering
	%		\includegraphics[width=0.4\textwidth]{graphs/MAP_1.png}
	%		\caption{MAP Scores}
	%		\label{RQ1-map-1}
	%	\end{minipage}
	%	\begin{minipage}{0.4\textwidth}
	%	\centering
	%	\includegraphics[width=0.4\textwidth]{graphs/nDCG_1.png}
%	\caption{Scores Changes from Top 1 to Top 100 on FFMPeg+Qemu Dataset}
%	\label{RQ1-score-3}
	%\end{minipage}
%\end{figure}

In Table~\ref{RQ1-data-3-top10}, among the top 10 prediction results,
{\tool} has the most correct predictions (6 vulnerable methods). The
vulnerable methods correctly detected by {\tool} are also pushed
higher in the top-10 ranked list with 4 correct results out of 5 top
results. All other baselines have only 0--1 correct detection in the
top-5 list. Importantly, the first rank for {\tool} (i.e., the rank of
the first correctly detected vulnerable methods) is 1st, while those
of the baselines are 4$^{th}$, 5$^{th}$, 5$^{th}$, 6$^{th}$, and
7$^{th}$ (the bold values in Table~\ref{RQ1-data-3-top10}). 
Moreover, \tool can detect 14, 35, and 64 vulnerabilities among top-20, top-50, and top-100 prediction results. 

%Due to the page limit, we calculate the MAP and nDCG scores for top-1 to top-100 lists and generate figures (Figure \ref{RQ1-score-1}) to show the performance of \tool and baselines.

%The table \ref{RQ1-data-3-top10} shows the prediction results for top
%10 on FFMPeg+Qemu Dataset. As we can see from the table, our approach
%can correctly predict the top-1 ranking results while other approaches
%cannot. The top ranking for the correctly predicted method for Reveal
%is 4 and for Devign and Russell et al. is 5. And for SySeVR, the top
%ranking is 6 while the VulDeePecker is even worse and the top ranking
%for it is only 7. As for the total number of the correctly predicted
%results, our approach can correctly predict 6 methods in the top-10
%predicted results while Reveal, Devign, Russell et al., SySeVR, and
%VulDeePecker can only fix 5, 4, 4, 4, and 3. These result proves that
%our approach can perform better on the top rankings.

Tables \ref{RQ1-data-3}, \ref{RQ1-data-1}, and \ref{RQ1-data-2} show
the comparison among the approaches on three datasets.  {\tool}
consistently performs better in all the metrics
(Table~\ref{RQ1-data-3}). For nDCG$@$\{1,3\}, all the baselines get zeros
because they did not have correct detections in top-3 results. {\tool}
can improve nDCG$@$10 from 43\%--84\% and nDCG$@$20 from 37\%--71\%
as compared to the baselines. Higher nDCG indicates that {\tool}
achieves the ranking closer to the perfect ranking and the
correct vulnerable methods appear higher in the top list.

For MAP scores, {\tool} relatively improves over the baselines from
105\%--255\% for top-10 and from 53\%--116\% for top 20. With higher
MAP, {\tool} has higher precision on average for all the top-ranked
positions in the top list. That is, the top-ranked result is highly
precise in detecting the vulnerable methods.

{\tool} also achieves better first ranking (FR) and average ranking (AR).
While its best FR is 1 and that of next best performer is 4. For
AR$@$10, a correct vulnerable method is on average ranked by {\tool}
2.7--4.0 positions higher in the ranked list than by the
baselines. Our tool also has relatively higher AUC from
6\%--24\%.

The comparative results on \code{Fan} and \code{Reveal} datasets are
similar (Tables~\ref{RQ1-data-1} and~\ref{RQ1-data-2}). In \code{Fan}
dataset, {\tool} can improve the nDCG and MAP
scores over the baselines by 26\%--43\%, 50\%--170\% for top-10, and 21\%--475\%,
40\%--250\% for top-20. {\tool}'s FRs and ARs are better from 2--6
positions and 0.7--2.7 positions for top 10, and 2--13 positions and
1.6--9.1 positions for top 20. In \code{Reveal} dataset, the
improvements in nDCG, MAP, FR, and AR are 33\%--73\%, 42\%--209\%,
2--7 positions, and 0--4 positions for top 10, and
19\%--111\%, 28\%--236\%, 2--12 positions, and 1.2--6.2 positions for
top 20.

The results on three datasets are different due to the ratio between
the vulnerable and non-vulnerable methods. That ratio is 1:16 and 1:9.9
in \code{Fan} and \code{Reveal} datasets. That number is 1:1.2
in \code{FFMPeg+Qemu} dataset, thus, there are more vulnerable methods,
and the results are consistently higher across all the models.

\begin{table}[t]
  \caption{RQ1. Method-Level VD on Fan Dataset}
  \vspace{-0.1in}
	\begin{center}
		\footnotesize
		\renewcommand{\arraystretch}{1}
		\newcommand{\tabincell}[2]{\begin{tabular}{@{}#1@{}}#2\end{tabular}}
		\begin{tabular}{p{1.1cm}<{\centering}|p{1cm}<{\centering}|p{0.7cm}<{\centering}|p{0.7cm}<{\centering}|p{0.7cm}<{\centering}|p{0.7cm}<{\centering}|p{1cm}<{\centering}}
			\hline
			                      &  VulDee- -Pecker  & SySeVR  & Russell {\em et al.}   & Devign     & Reveal        & {\tool}    \\
			\hline
						    nDCG@1&     0   &0&0&0&  0& 0\\
%			  		    	nDCG@3&     0   &0&0&0&0& 0\\
                            nDCG@5&     0   &0&0&0&0& 0.5\\
                           nDCG@10&     0   &0&0.30&0.33&0.34& 0.43\\
                           nDCG@15&     0   &0&0.28&0.30&0.37& 0.45\\
			  			   nDCG@20&        0.08&0.23&0.31&0.32&0.38& 0.46\\
			  			   \hline
			  			   MAP@1&       0   &0&0&0&0& 0\\
%			  			   MAP@3&       0   &0&0&0&0& 0\\
			  			   MAP@5&       0   &0&0&0&0& 0.25\\
			  			   MAP@10&      0   &0&0.1&0.13&0.18& 0.27\\
			  			   MAP@15&      0  &0&0.12&0.14&0.21& 0.28\\
			  			   MAP@20&      0.08  &0.24&0.14&0.15&0.20& 0.28\\
			  			   \hline
			  			   FR@1&      n/a  &n/a&n/a&n/a&n/a& n/a\\
%			  			   FR@3&      n/a  &n/a&n/a&n/a&n/a& n/a\\
			  			   FR@5&      n/a  &n/a&n/a&n/a&n/a& 4\\
			  			   FR@10&     n/a   &n/a&10&8&6& 4\\
			  			   FR@15&     n/a   &n/a&10&8&6& 4\\
			  			   FR@20&      19  &16&10&8&6& 4\\
			  			   \hline
			  			   AR@1&       n/a &n/a&n/a&n/a&n/a&n/a \\
%			  			   AR@3&       n/a &n/a&n/a&n/a&n/a&n/a \\
			  			   AR@5&       n/a &n/a&n/a&n/a&n/a& 4\\
			  			   AR@10&      n/a  &n/a&10&8&8& 7.3\\
			  			   AR@15&      n/a  &n/a&12&10.5&9.3& 8.5\\
			  			   AR@20&      19.5  &18&13.3&13.3&12& 10.4\\
			  			   \hline
			  			   AUC&        0.72&0.81&0.82&0.75&0.82& 0.9\\
			  					
			\hline
		\end{tabular}
		\label{RQ1-data-1}
	\end{center}
        \vspace{-0.04in}
\end{table}

\begin{table}[t]
  \caption{RQ1. Method-Level VD on Reveal Dataset}
  \vspace{-0.1in}
	\begin{center}
		\footnotesize
		\renewcommand{\arraystretch}{1}
		\begin{tabular}{p{1.1cm}<{\centering}|p{1cm}<{\centering}|p{0.7cm}<{\centering}|p{0.7cm}<{\centering}|p{0.7cm}<{\centering}|p{0.7cm}<{\centering}|p{1cm}<{\centering}}
			\hline
                     &  VulDee- -Pecker  & SySeVR  & Russell {\em et al.}   & Devign     & Reveal        & {\tool}    \\
			\hline
nDCG@1&      0  &0&0&0&0& 0\\
nDCG@3&      0  &0&0&0&0& 0.63\\
nDCG@5&      0  &0&0&0&0.43& 0.53\\
nDCG@10&     0   &0.30&0.32&0.34&0.39& 0.52\\
nDCG@15&     0.26   &0.28&0.32&0.39&0.42& 0.55\\
nDCG@20&     0.27   &0.33&0.35&0.43&0.48& 0.57\\
\hline
MAP@1&     0   &0&0&0&0& 0\\
MAP@3&     0   &0&0&0&0& 0.33\\
MAP@5&     0   &0&0&0&0.2& 0.37\\
MAP@10&    0    &0.11&0.11&0.18&0.24& 0.34\\
MAP@15&    0.07    &0.12&0.16&0.23&0.25& 0.36\\
MAP@20&    0.11    &0.15&0.18&0.36&0.29& 0.37\\
\hline
FR@1&       n/a &n/a&n/a&n/a&n/a& n/a\\
FR@3&      n/a  &n/a&n/a&n/a&n/a& 3\\
FR@5&       n/a &n/a&n/a&n/a&5& 3\\
FR@10&      n/a  &10&9&7&5&3 \\
FR@15&      15  &10&9&7&5& 3\\
FR@20&     15   &10&9&7&5& 3\\
\hline
AR@1&       n/a &n/a&n/a&n/a&n/a& n/a\\
AR@3&    n/a    &n/a&n/a&n/a&n/a& 3\\
AR@5&    n/a    &n/a&n/a&n/a&5& 4\\
AR@10&     n/a   &10&9&8&6&6 \\
AR@15&      15  &12.5&12&10.5&9.8& 9.5\\
AR@20&      18  &15.5&13.3&12.7&13& 11.8\\
\hline
AUC&        0.65&0.76&0.75&0.72&0.74& 0.81\\
			\hline
		\end{tabular}
		\label{RQ1-data-2}
	\end{center}
        \vspace{-0.07in}
\end{table}

\begin{table}[t]
	\caption{RQ1. Precision and Recall Results of Method-Level VD on Three Datasets (P: Precision; R: Recall; F: F score)}
	\vspace{-0.1in}
	\begin{center}
		\footnotesize
		\renewcommand{\arraystretch}{1}
		\begin{tabular}{p{1.4cm}<{\centering}|p{0.35cm}<{\centering}|p{0.35cm}<{\centering}|p{0.35cm}<{\centering}|p{0.35cm}<{\centering}|p{0.35cm}<{\centering}|p{0.35cm}<{\centering}|p{0.35cm}<{\centering}|p{0.35cm}<{\centering}|p{0.35cm}<{\centering}}
			\hline
			                &  \multicolumn{3}{c|}{FFMPeg+Qemu }  & \multicolumn{3}{c|}{Fan }     & \multicolumn{3}{c}{Reveal }    \\
			\hline
			                &  P  & R & F  & P   & R  &F   & P        & R   &F \\
			\hline
			VulDeePecker &0.49&0.27&0.35&0.12&0.49&0.19&0.19&0.14&0.17\\
			SySeVR&0.50&0.66&0.56&0.15&\textbf{0.74}&0.27&0.24&0.42&0.31\\
			Russell {\em et al.}&0.55&0.41&0.45&0.16&0.48&0.24&0.26&0.12&0.16\\
			Devign&0.52&0.63&0.57&0.18&0.52&0.26&0.33&0.32&0.32\\
			Reveal&0.55&\textbf{0.73}&0.62&0.19&\textbf{0.74}&0.30&0.31&\textbf{0.58}&0.40\\
			\hline
			{\tool} &\textbf{0.60}&0.72&\textbf{0.65}& \textbf{0.23} &0.72& \textbf{0.35}& \textbf{0.39}&0.52& \textbf{0.45} \\
			\hline
		\end{tabular}
		\label{RQ1-pr}
	\end{center}
	\vspace{-0.07in}
\end{table}

Table~\ref{RQ1-pr} shows the results of Precision and Recall of our {\tool} and the baselines.
%To compare the overall performance of the approaches, we use Precision and Recall scores to evaluate \tool and all baselines on all three datasets. 
%As shown in Table~\cite{RQ1-pr}, 
Specifically, \tool has higher precision than all the baselines on three datasets. \tool can improve the Precision by 2.6\%-105\%. 
For the Recall, \tool is marginally lower than Reveal on Fan and FFMPeg+Qemu datasets (i.e., 1.4\% and 2.7\%) and SySeVR on Fan dataset (i.e., 2.7\%). 
On the Reveal Dataset, \tool can improve Reveal by 25.8\% in terms of Precision, but decrease Recall by 10.3\%. However, in terms of F1 score, \tool can improve the best performed baseline Reveal by 4.8\% on FFMPeg+Qemu dataset, 16.7\% on Fan dataset, and 12.5\% on the Reveal Dataset.

%Except for these four conditions, \tool can improve all other baselines by 9\%-271\%. With such results, \tool shows a higher overall performance than all baselines on all three datasets.

\begin{figure}[t]
	\centering
	\subfloat[MAP Scores]{{\includegraphics[width=0.235\textwidth]{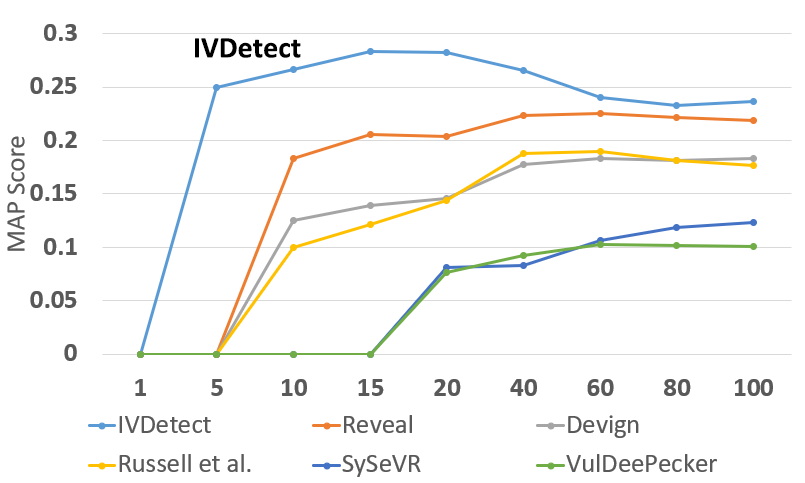}}}
	\subfloat[nDCG Scores]{{\includegraphics[width=0.235\textwidth]{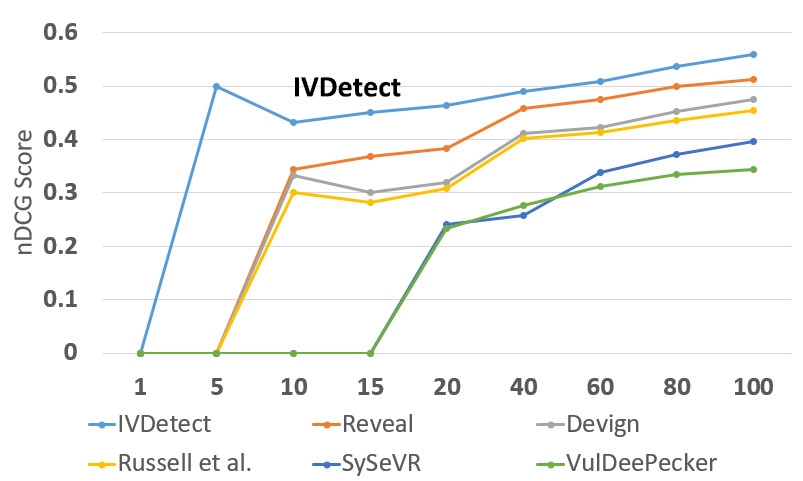}}}
        \vspace{-0.07in}
	\caption{Scores from Top 1 to Top 100 on Fan Dataset}
	\label{RQ1-score-1}
\end{figure}

Figure~\ref{RQ1-score-1} shows that {\tool} consistently has better
MAP and nDCG scores when considering top-1 to top-100 ranked lists.

\begin{figure}[t]
	\centering
	\subfloat[nDCG \& MAP Scores]{{\includegraphics[width=0.235\textwidth]{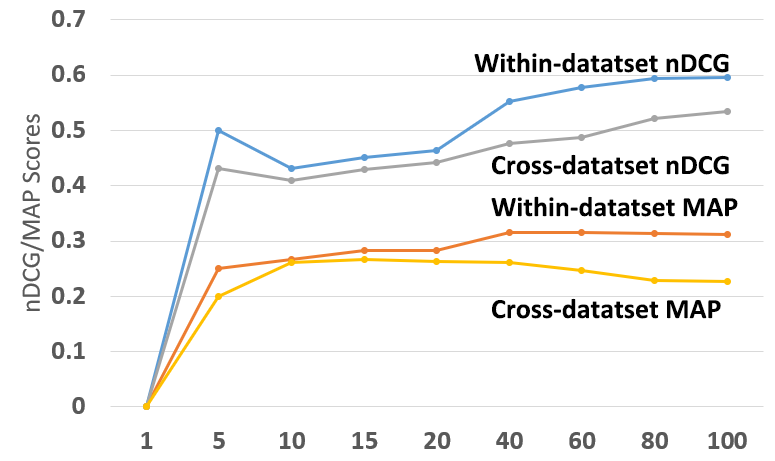}}}
	\subfloat[FR and AR]{{\includegraphics[width=0.235\textwidth]{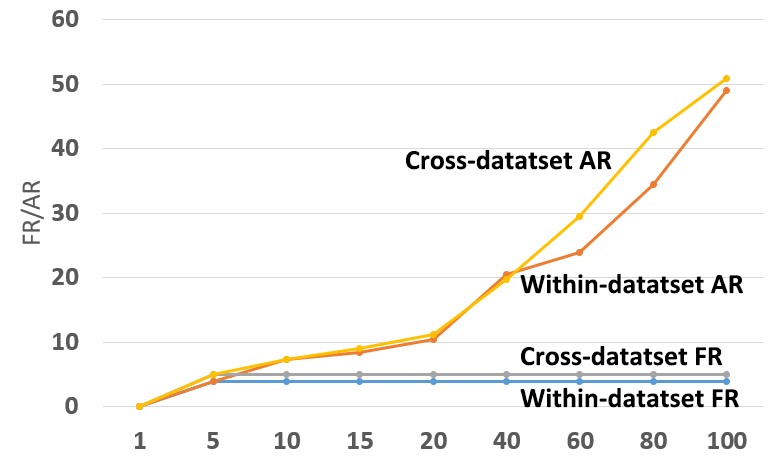}}}
        \vspace{-0.06in}
	\caption{RQ1. Cross-Dataset Validation: Training on Reveal and FFMPeg+Qemu Datasets, testing on Fan Dataset.}
	\label{RQ1-cross}
\end{figure}

For cross-dataset validation, as seen in Figure~\ref{RQ1-cross}, the
results for MAP and nDCG in within-dataset setting are better than
those in cross-dataset setting. This is expected because the model
might see similar vulnerable code before in the same projects in the
same dataset. The FR and AR values for cross-dataset setting are one
rank higher than those of within-dataset setting.

%As seen in Figure~\ref{RQ1-cross}, {\tool} achieves better results in the within-dataset setting than in the cross-dataset one. This is expected as the training and testing data is from the same dataset with the same projects in the within-dataset setting, thus a model may see similar vulnerabilities. But overall, both the cross-project setting and within-project setting can have a similar nDCG and MAP scores changing trend. And the differences in the scores are also not too much. As for the FR and AR, the cross-project setting has 1 ranking worse than the within the within-project setting and the AR lines are also very close. So the cross dataset setting will not influence the performance of our approach too much.

\begin{figure}[t]
	\centering
	\includegraphics[width=3.2in]{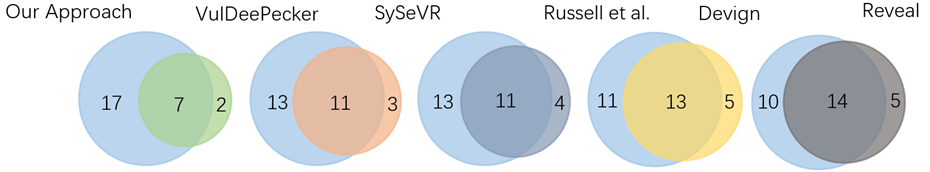}
        \vspace{-5pt}
	\caption{Overlapping Analysis}
	\label{RQ1-overlap}
\end{figure}

Figure~\ref{RQ1-overlap} shows our analysis on the overlapping results
between {\tool} and the baselines on \code{Fan} dataset for
top-100. As seen, {\tool} can detect 17, 13, 13, 11, and 10 vulnerable
methods that VulDeePecker, SySeVR, Russell, Devign, and Reveal missed,
respectively, while they can detect only 2,3, 4, 5, and 5 vulnerable
methods that {\tool} missed. In summary, {\tool} can detect 15, 10, 9,
6, and 5 more vulnerable methods than the baselines.

%% file: sections/rq2-results.tex
\subsection{{\bf RQ2. Comparison with other Interpretation Models for Fine-grained VD Interpretation}}

%\begin{table}[t]
%	\caption{RQ2. Fine-grained VD Interpretation Comparison}
%	\begin{center}
%		\small
%		\renewcommand{\arraystretch}{1} 
%		\begin{tabular}{p{1.5cm}<{\centering}|p{0.3cm}<{\centering}|p{0.3cm}<{\centering}|p{0.3cm}<{\centering}|p{0.3cm}<{\centering}|p{0.3cm}<{\centering}|p{0.3cm}<{\centering}|p{0.3cm}<{\centering}|p{0.3cm}<{\centering}|p{0.3cm}<{\centering}|p{0.3cm}<{\centering}|}
%			\hline
%		                	&  \multicolumn{10}{c|}{Accuracy}  \\
%		                	\cline{2-11}
%		                	           	&  N1 & N2 &  N3 & N4 &N5 & N6 &N7 & N8 &N9 & N10  \\
%			\hline
%			ATT                    & 0.01 & 0.16 & 0.41 & 0.54 & 0.59 & 0.60 & 0.62  & 0.63 & 0.64  &0.65 \\
%			GRAD                   &  0.01 & 0.19 & 0.43 & 0.54 & 0.59 & 0.62 & 0.63 & 0.65&0.66  & 0.67 \\
%			\hline
%			GNNExplainer          & 0.05 & 0.30 & 0.54 & 0.63 & 0.67 & 0.68 & 0.70 & 0.72&0.72  & 0.73 \\
%			\hline
%		\end{tabular}
%		Nx: x is the number of statements in the interpretation
%		\label{fig:RQ2}
%	\end{center}
%\end{table}

\begin{table}[t]
  \caption{RQ2. Fine-grained VD Interpretation Comparison}
  \vspace{-5pt}
	\begin{center}
		\small
		\renewcommand{\arraystretch}{1} 
		\begin{tabular}{p{0.6cm}<{\centering}|p{0.25cm}<{\centering}|p{0.25cm}<{\centering}|p{0.25cm}<{\centering}|p{0.25cm}<{\centering}|p{0.25cm}<{\centering}|p{0.26cm}<{\centering}|p{0.26cm}<{\centering}|p{0.26cm}<{\centering}|p{0.26cm}<{\centering}|p{0.26cm}<{\centering}|p{0.3cm}<{\centering}p{0.3cm}<{\centering}}
			\hline
		              Interp.  	&  \multicolumn{10}{c|}{Accuracy}  & \multirow{2}{*}{MFR}    & \multirow{2}{*}{MAR}       \\
		                	\cline{2-11}
		                	     Model      	&  N1 & N2 &  N3 & N4 &N5 & N6 &N7 & N8 &N9 & N10  &     &       \\
			\hline
			ATT                    & 0.01 & 0.16 & 0.41 & 0.54 & 0.59 & 0.60 & 0.62  & 0.63 & 0.64  &0.65 &  4.8   &  6.3    \\
			GRAD                   &  0.01 & 0.19 & 0.43 & 0.54 & 0.59 & 0.62 & 0.63 & 0.65&0.66  & 0.67 &  4.2   &  5.6    \\
			\hline
			GE          & 0.05 & 0.30 & 0.54 & 0.63 & 0.67 & 0.68 & 0.70 & 0.72&0.72  & 0.73  &  3.5   &  5.0    \\
			\hline
		\end{tabular}
		GE: GNNExplainer; Nx: x is the number of nodes in the interpretation
		\label{fig:RQ2}
	\end{center}
\end{table}

Table~\ref{fig:RQ2} shows the accuracy of different interpretation
models. As seen, using GNNExplainer improves over ATT and GRAD from
12.3\%--400\% and 9.0\%--400\% in accuracy, respectively, as we vary
the size of interpretation sub-graphs (i.e., the number of
statements) from 1--10. Higher accuracy indicates that {\tool} can
provide better fine-grained vulnerability detection interpretation at
the statement level. That is, {\em in more cases, if {\tool} detects
correctly vulnerable methods, it can point out more precisely the
vulnerable statements} relevant to the vulnerabilities. For ranking
vulnerable statements, using GNNExplainer improves MFR by 0.7 and 1.3
ranks, and improves MAR by 0.6 and 1.3 ranks over ATT and GRAD.

%As seen, comparing with \emph{ATT}, using GNNExplainer can improve the
%accuracy by 12.3\%--400\%, and improve MFR by 0.7 and 1.3 and the MAR
%can be improved by 0.6 and 1.3 respectively.
%
%that using GNNExplainer in the interpretation section of our approach
%can perform better than the other two baselines on fine-grained
%vulnerability detection. Specifically, comparing \emph{ATT}
%with \emph{GNNExplainer}, using GNNExplainer can improve the accuracy
%by 12.3\%--400\% and 9.0\%--400\% respectively while the MFR can be
%improved by 0.7 and 1.3 and the MAR can be improved by 0.6 and 1.3
%respectively.

ATT uses the edge attention in the Graph Attention Network to assign
the weights for the edges, while GNNExplainer directly gives a score
for the subgraph after masking. Thus, for the case in which there are
more than one path from a node to another, the weight for an edge is
the average weight of the weights through multiple paths, i.e., ATT might
be less precise than GNNExplainer. GRAD computes the
gradient of the loss function with respect to the input for computing
the weight of an edge. However, such gradient-based approach may not
perform well with respect to the discrete inputs (an input graph is
represented as an adjacency matrix).

%GNNExplainer can have higher performance than the ATT because even the ATT directly uses the edge attention in GAT to calculate the importance of each edge, but it is not clear which attention weight is the best to use. For example, node $a$ is connected to node $b$. When dealing with node $b$, the node $a$ can be directly linked to $b$, but also can be linked by $c$ and then to $B$. In this case, in ATT, each edge’s importance is computed as the average attention weight across all situations while the GNNExpalainer could directly give one importance score without considering this.

%Comparing \emph{GRAD} with \emph{GNNExplainer}, using GNNExplainer can improve the accuracy by 6.7\% respectively and can improve the MFR and MAR by 0.7 and 0.6.

%GRAD computes the gradient of the GNN’s loss function with respect to the adjacency matrix to evaluate the importance of the edges. But the gradient-based approaches often have the problem that they often may not perform well with discrete inputs which are very common in adjacency matrix (binary matrix) in GNN \ref{GNNExplainer}, GRAD is also in the case. This is the reason why the GNNExplainer can perform better in our problem than GRAD in this RQ.

As the number of nodes in $\mathcal{G}_M$ increases, the number of
statements covered also increases, accuracy is higher. However, the
computation time is higher and developers need to investigate more
statements. As seen, when the number of statements is higher than 5,
accuracy increases more slowly. Thus, we chose 5 as a default.

%% file: sections/rq4-results.tex
\subsection{{\bf RQ3. Vulnerable Code Pattern Analysis}}
\label{pattern:sec}

%{\color{red}{I have missed some experiments on the level of dependencies in detection. I am doing the experiments now, will add here soon.}}

\begin{table}[t]
	\caption{RQ3. Numbers of Vulnerable Code Patterns}
        \vspace{-0.06in}
        \small
	\begin{center}
		\small
		\renewcommand{\arraystretch}{1} 
		\begin{tabular}{p{1.5cm}<{\centering}|p{0.75cm}<{\centering}|p{0.75cm}<{\centering}|p{0.75cm}<{\centering}|p{0.75cm}<{\centering}}
			\hline
			                & thres=2   & thres=3   & thres=4   & thres=5   \\           			\hline
			size=2          & 47        & 36        & 22        & 7         \\
			size=3          & 25        & 27        & 19        & 6         \\
			size=4          & 23        & 22        & 16        & 5         \\
			size=5          & 22        & 21        & 11        & 2         \\
			\hline
			Total           & 117       & 102       & 68        & 21        \\
			\hline
		\end{tabular}
		\label{pattern:tab}
		
	\end{center}
        \vspace{-0.06in}
\end{table}

%\begin{table}[t]
%	\caption{RQ3. Numbers of Vulnerable Code Patterns}
%        \vspace{-0.06in}
%	\begin{center}
%		\small
%		\renewcommand{\arraystretch}{1} 
%		\begin{tabular}{p{1.5cm}<{\centering}|p{0.75cm}<{\centering}|p{0.75cm}<{\centering}|p{0.75cm}<{\centering}|p{0.75cm}<{\centering}}
%			\hline
%			                & freq=2   & freq=3   & freq=4   & freq=5   \\           			\hline
%			maxsize=2                      &  &  &  & \\
%			maxsize=3                      &  &  &  &  \\
%			maxsize=4                  &  &  &  &  \\
%			maxsize=5                  &  &  &  &  \\
%			\hline
%		\end{tabular}
%		\label{pattern}
		
%	\end{center}
%        \vspace{-0.06in}
%\end{table}

%                        \hline
%			Accuracy                & F2   & F3   & F4   &F5   \\           			\hline
%			S2                      &8.4\% &14.8\%&9.7\%&6.0\%\\
%			S3                      &14.6\%&18.5\%&14.6\%&9.4\%\\
%			S4                      &17.3\%&19.8\%&16.9\%&11.1\%\\
%			S5                      &19.7\%&21.3\%&18.2\%&13.9\%\\
%			\hline

%F: the frequency limit for the pattern (the pattern should at least
%appear the frequency limitation times); S: the max pattern size limit
%(the pattern should not be bigger than this); Accuracy: when the
%pattern has a least one statement overlapping with the statements that
%need to be fixed, we regard this pattern as correctly generated
%pattern.

This section describes another experiment that we exploit {\tool}'s
capability of providing interpretation sub-graphs to mine the patterns
of vulnerable code. A vulnerable code pattern is a fragment of
vulnerable code that repeats frequently, i.e., more than a certain
threshold. 
The detected vulnerability patterns and corresponding fixes can be the good sources for developers to learn about the vulnerable code that others have frequently made, and learn to fix vulnerable code in the same patterns.

From the results in RQ2, we first collected into a set
$\mathcal{G}$ the interpretation sub-graphs $\mathcal{G}_M$s with the
correctly detected statements as relevant to the vulnerability in the
methods. In total, we obtain +700 $\mathcal{G}_M$s. Note that
$\mathcal{G}_M$ is a sub-graph of PDG. For each $\mathcal{G}_M$, we
abstract out the variables' names with a keyword \code{VAR}, and the
literals with their data types. We then ran the sub-graph pattern
mining algorithm~\cite{fse09} on $\mathcal{G}$ with different
thresholds of frequencies and collected different sizes of the
sub-graph patterns. The outputs are the frequent isomorphic sub-graphs
within $\mathcal{G}_M$s, which are considered as vulnerable code
patterns because we chose $\mathcal{G}_M$ that contains correct
  interpretation statements relevant to the correctly detected
  vulnerabilities. After manual verification, we obtain
  a number of correct patterns (Table~\ref{pattern:tab}).
% Table~\ref{pattern:tab} shows the correct patterns
% after our manual verification.
  As seen, as the frequency threshold or the size of pattern is
  larger, the number of patterns decreases as expected. When they are
  both larger than 5, we found no pattern. Let us explain a few
  examples.

\begin{figure}[t]
	\centering
	\renewcommand{\lstlistingname}{Method}
	\lstset{
		numbers=left,
		numberstyle= \tiny,
		keywordstyle= \color{blue!70},
		commentstyle= \color{red!50!green!50!blue!50},
		frame=shadowbox,
		rulesepcolor= \color{red!20!green!20!blue!20} ,
		xleftmargin=2em,xrightmargin=1em, aboveskip=1em,
		framexleftmargin=0.5em,
		language=Java,
		basicstyle=\tiny\ttfamily,
                moredelim=**[is][\color{red}]{@}{@},
		escapeinside= {(*@}{@*)}
	}
\begin{minipage}[c]{0.45\textwidth}
	\begin{lstlisting}
// ========================PATTERN 1 =======================================        
	(*@{\color{red}{if (is$\_$link(STRINGLITERAL))}@*) {
		fprintf(stderr, "Error: invalid /etc/skel/.zshrc file\n"); // not in pattern
		(*@{\color{red}{exit(INTLITERAL);}@*)
	}
	(*@{\color{red}{if (copy$\_$file(STRINGLITERAL, VAR) == INTLITERAL)}@*) { ...
// ========================PATTERN 2 =======================================
	(*@{\color{orange}{VAR = udf$\_$get$\_$filename(VAR, VAR, VAR, VAR);}@*)
        (*@{\color{orange}{if (VAR $\&\&$ ...) goto LABEL;}@*)
	\end{lstlisting}
        \vspace{-0.18in}
	\caption{Vulnerable Code Patterns}
        \vspace{-0.12in}
		\label{fig:example_vul}
\end{minipage}
\end{figure}

%(*@{\color{red}{if (VAR $\&\&$ ...) goto LABEL;}@*)

Figure~\ref{fig:example_vul} shows two examples of vulnerable code
patterns. The first pattern (lines 2,4, and 6) shows an API misuse in
the project \code{firejail} involving \code{is$\_$link(...)},
\code{exit}, and \code{copy$\_$file(...)}. The usage is to check the
validity of a link, and if yes to copy the file, or otherwise to stop
the execution. This pattern appeared three times with different string
literals and was fixed by developers to replace the statements. An
interesting observation is that {\tool} is able to eliminate the
\code{fprintf} statement at line 2 from the interpretation sub-graph,
thus, eliminating it from the pattern, even though the \code{fprintf}
statement appears with the other statements three times in the
project. This shows a benefit of {\tool} because if a tool does not have
statement-level VD interpretation and it mines pattern from the entire methods, it
will incorrectly include \code{fprintf} in the pattern. The second
pattern (lines 8--9) shows a pattern involving a vulnerable method
call \code{udf$\_$get$\_$filename}, and the checking on its return
value. The method later was fixed to add the 5$^{th}$ parameter.

\begin{figure}[t]
	\centering
	\renewcommand{\lstlistingname}{Method}
	\lstset{
		numbers=left,
		numberstyle= \tiny,
		keywordstyle= \color{blue!70},
		commentstyle= \color{red!50!green!50!blue!50},
		frame=shadowbox,
		rulesepcolor= \color{red!20!green!20!blue!20} ,
		xleftmargin=2em,xrightmargin=1em, aboveskip=1em,
		framexleftmargin=0.5em,
		language=Java,
		basicstyle=\tiny\ttfamily,
		moredelim=**[is][\color{red}]{@}{@},
		escapeinside= {(*@}{@*)},
	}
	\begin{minipage}[c]{0.45\textwidth}
		\begin{lstlisting}
// ===================== FIXING PATTERN 1 =========================
(*@{\color{red}{-	VAR = fl6$\_$update$\_$dst(VAR, VAR, VAR);}@*)
(*@{\color{blue}{+	rcu$\_$read$\_$lock();}@*)
(*@{\color{blue}{+	final$\_$p = fl6$\_$update$\_$dst(VAR, rcu$\_$dereference(VAR), VAR);}@*)
(*@{\color{blue}{+	rcu$\_$read$\_$unlock();}@*)
// ===================== FIXING PATTERN 2 =========================
(*@{\color{red}{-	char VAR = malloc (VAR);}@*)
(*@{\color{blue}{+   char VAR;}@*)
(*@{\color{blue}{+   if (VAR < 0 || VAR > LITCONST) \{ }@*)
(*@{\color{blue}{+ \quad      error$\_$line (STRINGLITERAL, VAR);}@*)
(*@{\color{blue}{+ \quad      return LITCONST;}@*)
(*@{\color{blue}{+   \} }@*)	
(*@{\color{blue}{+       VAR = malloc (VAR);}@*)
		\end{lstlisting}
                \vspace{-0.18in}
		\caption{Fixing Patterns (-: removal, +: addition)}
                \vspace{-0.12in}
                \label{fig:fixpattern}
	\end{minipage}
\end{figure}

Another interesting finding is that {\tool} enables the discovery of
not only vulnerable code patterns but also the fixing patterns for
them. Figure~\ref{fig:fixpattern} shows two fixing patterns for
vulnerable code. The first vulnerability (from Linux kernel), lines
2--5, is about the method \code{f16$\_$update$\_$dst}(...). According
to the commit log, to avoid another thread changing a data record
concurrently, developers need to provide mutual exclusion access and
deferencing. This fixing pattern was repeated 3 times in the methods
\code{dccp$\_$v6$\_$send$\_$response},
\code{inet6$\_$csk$\_$route$\_$req}, and
\code{net6$\_$csk$\_$route$\_$socket}.  This fixing pattern would be
useful for a developer to learn the fix from one method and apply to
the other two methods. The second pattern (lines 7--13) shows a fixing
pattern to a vulnerability on buffer overflow with the \code{malloc}
call in \code{ParseDsdiffHeaderConfig} method of WavPack
5.0. According to CVE-2018-7253, this problem {\em ``allows a remote
  attacker to cause a denial-of-service (heap-based buffer over-read)
  or possibly overwrite the heap via a maliciously crafted DSDIFF
  file''}.  This fixing pattern occurred three times in the same
project.

%% file: sections/rq3-results.tex
\subsection{{\bf RQ4. Sensitivity Analysis for Features}}

%\begin{table}[t]
%	\caption{RQ3. Evaluation for the Impact of Internal Features.}
%	\begin{center}
%		\small
%		\renewcommand{\arraystretch}{1} 
%		\begin{tabular}{p{1.5cm}|p{0.9cm}<{\centering}|p{0.8cm}<{\centering}|p{0.9cm}<{\centering}|p{0.9cm}<{\centering}|p{0.5cm}<{\centering}%%|p{0.5cm}<{\centering}}
%			\hline
%		                                	&  Precision & Recall & F-score     & Accuracy     & AUC        & FPR    \\
%			\hline
%			ST (A)                          &  0.16      &  0.60  &   0.25      &   0.75       & 0.75       & 0.26   \\
%			(A)+SST (B)                     &  0.16      &  0.61  &   0.26      &   0.77       & 0.79       & 0.24   \\
%			(B)+SA (C)                      &  0.17      &  0.63  &   0.27      &   0.80       & 0.78       & 0.23   \\
%			(C)+V (D)                       &  0.18      &  0.64  &   0.28      &   0.82       & 0.82       & 0.21   \\
%	        (D)+CD (E)                      &  0.21      &  0.65  &   0.32      &   0.87       & 0.84       & 0.18   \\
%			(E)+DD (F)                      &  0.23      &  0.72  &   0.35      &   0.92       & 0.87       & 0.16   \\
%			\hline
%		\end{tabular}
	%	\label{RQ3}
%		ST: sequence of tokens; SST: sequence of sub-tokens; SA: sub-AST; V: variables; CD: control dependencies; DD: data dependencies
%	\end{center}
%\end{table}

Table~\ref{sensi:tab} shows the changes to the metrics as we
incrementally added each internal feature into our model in
Figure~\ref{fig:feature}. Generally, each internal feature contributes
positively to the better performance of {\tool}, as both the score metrics
(nDCG, MAP, and AUC) and the ranking metrics (FR and AR)
are improved.

When {\tool} considers only the sequence of tokens (ST) in the code,
the first correct detection (FR) is at the position 14, thus,
nDCG$@$\{1,5,10\}=0 and MAP$@$\{1,5,10\}=0 (not shown). When
considering the code as the sequence of sub-tokens (SST), {\tool}
deals with the unique tokens better because the sub-tokens appear more
frequently than the tokens~\cite{icse20-methodname}. At top-20, FR
improves 2 positions, AR improves 4.5 positions, and nDCG and MAP
relatively improve 3.8\% and 22.2\%. When AST is additionally
considered, the model can distinguish vulnerable code structures and
statements. At top-20, FR and AR improve 1 and 1.5 positions, and nDCG
and MAP improve 7.4\% and 18.1\%. However, FR is still 11 and
nDCG$@$\{1,5,10\}=0 and MAP$@$\{1,5,10\}=0 (not shown), because tokens
and AST do not help much discriminate the vulnerable statements.

\begin{table}[t]
	\caption{RQ4. Evaluation for the Impact of Internal Features.}
        \vspace{-0.08in}
	\begin{center}
		\footnotesize
		\renewcommand{\arraystretch}{1} 
		\newcommand{\tabincell}[2]{\begin{tabular}{@{}#1@{}}#2\end{tabular}}
		\begin{tabular}{p{1.2cm}<{\centering}|p{0.5cm}<{\centering}|p{0.8cm}<{\centering}|p{0.8cm}<{\centering}|p{0.8cm}<{\centering}|p{0.8cm}<{\centering}|p{0.8cm}<{\centering}}
			\hline
			&  {\bf ST} (A)   & (A)+{\bf SST} (B)  & (B)+{\bf AST} (C)   & (C)+{\bf Var} (D)     & (D)+{\bf CD} (E)        & (E)+{\bf DD} (F)     \\
			\hline
%			nDCG@1&     0   &0&0&0&  0& 0\\
%			nDCG@5&     0   &0&0&0&0.43& 0.50\\
%			nDCG@10&     0   &0&0&0.33&0.40& 0.43\\
			nDCG@15&     0.25   &0.27&0.29&0.35&0.42& 0.45\\
			nDCG@20&        0.26&0.27&0.29&0.37&0.44& 0.46\\
			\hline
%			MAP@1&       0   &0&0&0&0& 0\\
%			MAP@5&       0   &0&0&0&0.2& 0.25\\
%			MAP@10&      0   &0&0&0.18&0.24& 0.27\\
			MAP@15&      0.07  &0.11&0.12&0.19&0.26& 0.28\\
			MAP@20&      0.09  &0.11&0.13&0.19&0.26& 0.28\\
			\hline
%			FR@1&      n/a  &n/a&n/a&n/a&n/a& n/a\\
%			FR@5&      n/a  &n/a&n/a&n/a&5& 4\\
%			FR@10&     n/a   &n/a&n/a&7&5& 4\\
			FR@15&     14   &12&11&7&5& 4\\
			FR@20&     14  &12&11&7&5& 4\\
			\hline
%			AR@1&       n/a &n/a&n/a&n/a&n/a&n/a \\
%			AR@5&       n/a &n/a&n/a&n/a&6& 4\\
%			AR@10&      n/a  &n/a&n/a&8&8& 7.3\\
			AR@15&      14  &13.5&11&10.3&9& 8.5\\
			AR@20&      19.5  &15&13.5&12.5&11.2& 10.4\\
			\hline
			AUC&        0.75&0.76&0.77&0.83&0.85& 0.9\\
			
			\hline
		\end{tabular}
		\label{sensi:tab}
		{\bf ST}: sequence of tokens; {\bf SST}: sequence of sub-tokens; {\bf AST}: sub-AST; {\bf Var}: variables; {\bf CD}: control dependencies; {\bf DD}: data dependencies; {\bf F = {\tool}}
	\end{center}
\vspace{-0.12in}                        
\end{table}

The feature on variables also helps improve FR and AR from 11 to 7 and
13.5 to 12.5, and nDCG and MAP relatively improve 27.6\% and 46.2\% at
top 20. nDCG$@$10 and MAP$@$10 improve from 0 to 0.33 and to 0.18,
respectively (not shown). This feature allows the model to detect
similar incorrect variable usages.
By additionally integrating control dependencies (CD), FR and AR improve
from 7 down to 5 and 12.5 down to 11.2, and nDCG and MAP relatively
improve 18.9\% and 36.8\%. By adding data dependencies (DD), FR and AR
improve from 5 to 4 and 11.2 to 10.4. nDCG and MAP improve 4.5\% and
7.7\% for top 20. This result confirms that vulnerable code often
involves the statements with control and/or data
dependencies~\cite{zhou2019devign,chakraborty2020deep}.

\begin{figure}[t]
	\centering
	\renewcommand{\lstlistingname}{Method}
	\lstset{
		numbers=left,
		numberstyle= \tiny,
		keywordstyle= \color{blue!70},
		commentstyle= \color{red!50!green!50!blue!50},
		frame=shadowbox,
		rulesepcolor= \color{red!20!green!20!blue!20} ,
		xleftmargin=2em,xrightmargin=1em, aboveskip=1em,
		framexleftmargin=0.5em,
		language=Java,
		basicstyle=\tiny\ttfamily,
		moredelim=**[is][\color{red}]{@}{@},
		escapeinside= {(*@}{@*)},
	}
	\begin{lstlisting}
static int validate_group(struct perf_event *event)
{       ...
-	if (!validate_event(&fake_pmu, leader))
+	if (!validate_event(event->pmu, &fake_pmu, leader))
		return -EINVAL;

	list_for_each_entry(sibling, &leader->sibling_list, group_entry) {
-		if (!validate_event(&fake_pmu, sibling))
+		if (!validate_event(event->pmu, &fake_pmu, sibling))
			return -EINVAL;
	}	

-	if (!validate_event(&fake_pmu, event))
+	if (!validate_event(event->pmu, &fake_pmu, event))
		return -EINVAL; ...
}
	\end{lstlisting}
        \vspace{-0.16in}
	\caption{A Detected Vulnerable Method in Android kernel}
	\label{fig:example_2}
        \vspace{-0.06in}
\end{figure}

Figure~\ref{fig:example_2} shows a detected vulnerable method:
\code{validate$\_$event(...)} was vulnerable and replaced with a new
version with an additional parameter. We used the models (A)--(F) for
detection, and observed that the rank for
\code{validate$\_$event(...)} in the candidate list improves from
$140$ (A), to $121$ (B), $99$ (C), $71$ (D), $48$ (E), and $19$ (F).
While the features on tokens, sub-tokens, and AST are contributing,
they do not help much because the model did not see them in vulnerable
methods before. However, the variable/method names, especially
control/data dependencies between the surrounding statements and
\code{validate$\_$event(...)} help discriminate this vulnerability,
and push it to the top-20 list. Control dependencies (e.g., between
\code{validate$\_$event(...)}  and \code{return -EINVAL}) help improve
29 ranks. Generally, the improvement in ranking shows the positive
contributions of all the features.

This example also shows a fixing pattern appearing three times with
different variables \code{leader}, \code{sibling}, and \code{event}.

%% file: sections/rq5-results.tex
\subsection{{\bf RQ5. Sensitivity Analysis on Training Data}}

%\begin{table}[t]
%	\caption{RQ5. Evaluation for the Impact of External Features.}
%	\begin{center}
%		\small
%		\renewcommand{\arraystretch}{1} 
%		\begin{tabular}{p{1.7cm}|p{0.9cm}<{\centering}|p{0.8cm}<{\centering}|p{0.85cm}<{\centering}|p{0.9cm}<{\centering}|p{0.4cm}<{\c%entering}|p{0.4cm}<{\centering}}
%%			\hline
%			Train/Tune/Test                 &  Precision & Recall & F-score     & Accuracy     & AUC        & FPR    \\
%			\hline
%			40\%/30\%/30\%                  &  0.14      &  0.41  &   0.21      &   0.72       & 0.68       & 0.37   \\
%			50\%/25\%/25\%                  &  0.17      &  0.52  &   0.26      &   0.79       & 0.73       & 0.25   \\
%			60\%/20\%/20\%                  &  0.20      &  0.65  &   0.31      &   0.88       & 0.82       & 0.19   \\
%			70\%/15\%/15\%                  &  0.22      &  0.69  &   0.34      &   0.90       & 0.86       & 0.17   \\
%			80\%/10\%/10\%                  &  0.23      &  0.72  &   0.35      &   0.92       & 0.87       & 0.16   \\
%			\hline
%		\end{tabular}
%		\label{RQ5}
%	\end{center}
%\end{table}

\begin{table}[t]
  \caption{RQ5. Sensitivity Analysis on Training Data}
 \vspace{-0.1in}
\begin{center}
\footnotesize
\renewcommand{\arraystretch}{1}
\begin{tabular}{p{1.7cm}|p{0.9cm}<{\centering}|p{0.9cm}<{\centering}|p{0.85cm}<{\centering}|p{0.9cm}<{\centering}|p{0.4cm}<{\centering}}
\hline
Train/Tune/Test                 &  nDCG@20 & MAP@20 &   FR@20     & AR@20        & AUC \\
\hline
40\%/30\%/30\%                  &  0.26    &  0.09  &   12        &   15.5       & 0.69    \\
50\%/25\%/25\%                  &  0.33    &  0.16  &   8         &   12.3       & 0.74    \\
60\%/20\%/20\%                  &  0.43    &  0.25  &   5         &   11.6       & 0.85   \\
70\%/15\%/15\%                  &  0.44    &  0.26  &   5         &   11.2       & 0.87    \\
80\%/10\%/10\%                  &  0.46    &  0.28  &   4         &   10.4       & 0.9    \\
\hline
\end{tabular}
\label{RQ5}
\end{center}
%\vspace{-0.12in}
\end{table}

%We conducted experiments with different data splitting ratios and
%measured the performance. These splitting schemes include
%40\%/30\%/30\%, 50\%/25\%/25\%, 60\%/20\%/20\%, 70\%/15\%/15\%, and
%80\%/10\%/10\% for training, tuning, and testing sets. We used Fan
%dataset for this.
As seen in Table~\ref{RQ5}, with more training data, the performance
is better as expected. Even with 60\%/20\%/20\%, {\tool} still
achieves nCDG of 0.43 and MAP of 0.25, which are still higher than
those of the other baselines for top 20 (highest nDCG and MAP of the
baselines are 0.38 and 0.20). With 20\% less training data (60\% vs 80\%), {\tool} only drops AUC by 5.5\%. 

%Our model still performs well with a bit less training data (60\% instead of 80). 

%To analysis is the data splitting influence the model performance, we do the evaluation on different data splitting schemes on Fan et al. dataset. These splitting schemes include 40\%/30\%/30\%, 50\%/25\%/25\%, 60\%/20\%/20\%, 70\%/15\%/15\%, and 80\%/10\%/10\% for training/tuning/testing dataset. The results show that when we use more data to train the model and fewer data to do the prediction, the results would be better. But even by using 60\%/20\%/20\% setting, {\tool} still has 0.40 and 0.23 for nDCG and MAP scores. They are still higher than other baselines on top 20 (highest nDCG for baselines on top 20: 0.38; highest MAP for baselines on top 20: 0.20;). And the top ranking and average ranking in this stage are 5 and 11.2 which are also better than baselines on top 20 (best first ranking for baselines: 6; best average ranking for baselines: 12). As a summary for all of these, even though more training data can make our model learn more and perform better, but less training data does not influence our model performance too much. It means that our approach does not have the over fitting problem for the current setting.

\vspace{-0.04in}
\subsubsection*{{\bf Time Complexity}}
To generate the interpretation sub-graphs for all methods, it takes
about 9 days, 2 days, and 3 days to finish on Fan, Reveal, and
FFMPeg+Qemu datasets, respectively. It took 23, 7, 10 hours to train
{\tool} on Fan, Reveal, and FFMPeg+Qemu datasets. For
VD prediction, it takes only 1-2s per method.

%In table \ref{RQ5}, it shows that different data splitting schemes for training, tuning, and testing data. The results show that when we use more data to train the model and fewer data to do the prediction, the results would be better. But even by using 60\%/20\%/20\% setting, {\tool} still has 0.31, 0.83, 0.82, and 0.19 on F-score, Accuracy, AUC, and FPR, which is still higher than all other baselines on F-score (0.30 highest). AUC and FPR in this setting are the same as the best baselines while the Accuracy is 1\% lower than the best baselines. As a summary for all of these, our approach still can be regarded as better than other baselines in the 60\%/20\%/20\% setting.

%% file: sections/rq6-results.tex
%\subsubsection{{\bf RQ6. Time Complexity Analysis.}}

%{\color{red}{will add results here}}

%% file: sections/threats.tex
\vspace{-0.05in}
\subsubsection*{\bf Threats to Validity}
We only tested on the vulnerabilities in C and C++ code.  In
principle, {\tool} can apply to other programming languages.  We
tried our best to tune the baselines on same dataset for
fair~comparisons. We focus only on DL-based VD models.

%% file: sections/relat.tex
\vspace{-0.05in}
\section{Related Work}
%Here, we summarize the relevant existing vulnerability detection approaches and some achievements in the interpretable AI.

%\vspace{2pt}
%\noindent{\bf Rule-based and Machine learning-based Approaches.}

Various techniques have been developed to detect vulnerabilities.
%Security experts or researchers can utilize the fuzzing techniques~\cite{bohme2017coverage,wang2010taintscope,wen2020memlock,wang2017skyfire},
%symbolic execution~\cite{stephens2016driller,cha2015program,babic2011statically} or even manual auditing to hunt the vulnerabilities. However, the above approaches still require a large amount of time and resources to hunt the vulnerable code, even worse, most of the effort is wasted in analyzing the non-vulnerable code. Therefore, directly analyzing source code to identify the potential vulnerable code to help the vulnerability hunt is attracting a lot of attention.
%fuzzing (e.g., [9, 11, 39, 52, 63, 64, 67, 69, 70]), symbolic execution (e.g., [6, 10, 24, 60]) or manual auditing.
The rule-based approaches were developed to leverage known vulnerability patterns to discover possible vulnerable code, such as FlawFinder~\cite{FlawFinder}, RATS~\cite{RATS}, ITS4~\cite{viega2000its4}, Checkmarx~\cite{Checkmarx}, Fortify~\cite{HPFortify} and Coverity~\cite{Coverity}.
Typically, the patterns are manually defined by human experts. The state-of-the-art vulnerability detection tools using static analysis provide the
rules for each vulnerability type.
%For example,
%This category includes the lightweight methods that
%generate patterns from source code (e.g.,
%the well-known open source tools include FlawFinder~\cite{FlawFinder}, RATS~\cite{RATS}, ITS4~\cite{viega2000its4}, and the commercial tools include Checkmarx~\cite{Checkmarx}, Fortify~\cite{HPFortify} and Coverity~\cite{Coverity}.
%However, the rule-based approaches suffer the problem of high-false positives or high-false negatives and also more focus on one or few types of vulnerabilities.
%and the more comprehensive methods that
%generate patterns based on intermediate code (e.g., commercial
%tools Fortify~\cite{HPFortify} and Coverity~\cite{Coverity}).
%These tools can report
%the locations and the types of vulnerabilities, but cannot
%accurately distinguish between various vulnerable code and
%non-vulnerable code, resulting in high false positives or high
%false negatives~\cite{li2018vuldeepecker}.

Another type is machine learning (ML)-based or metrics-based.
Typically, these approaches require the human-crafted or summarized metrics as features to characterize vulnerabilities and train machine learning models on the defined features to predict whether a given code is vulnerable or not.
%Traditional machine learning-based methods rely on human
%experts for defining features to characterize vulnerabilities,and use traditional machine learning techniques [37], such as k-nearest neighbor and support vector machine, to classify vulnerable code and non-vulnerable code.
%For example,
%Ghaffarian and Shahriari~\cite{ghaffarian2017software} [38] reviewed the approaches of vulnerability analysis and discovery using machine-learning techniques, most of which are traditional machine learning-based methods.
Various ML-based approaches have been built on top of distinct metrics, such as terms and their occurrence frequencies~\cite{scandariato2014predicting}, imports and function calls~\cite{neuhaus2007predicting}, complexity, code churn, and developer activity~\cite{shin2010evaluating}, dependency relation ~\cite{neuhaus2009beauty}, API symbols and subtrees~\cite{yamaguchi2012generalized, yamaguchi2011vulnerability}.

Recently, deep learning (DL) has been applied to detect
vulnerabilities.  For example, some approaches train a DL model on
different code representations to detect vulnerabilities, such as the
lexical representations of functions in a synthetic
codebase~\cite{harer2018learning}, code snippets related to
API calls to detect two types of
vulnerabilities~\cite{li2018vuldeepecker}, syntax-based,
semantics-based, and vector representations~\cite{li2018sysevr},
graph-based representations~\cite{zhou2019devign}.
None of them is designed to provide interpretations for a model in term of
vulnerable statements.

%Our {\tool} is different from all of the above approaches.

%The main goal of {\tool} is to add more intelligence assistance (IA)
%using a small graph of code statements, key variables, and CWE
%description to explain why a detection model reaches a prediction.

%Nevertheless, there is no systematic comparative study to show the quantitative
%impact of different factors on the effectiveness of vulnerability
%detection. The present study follows VulDeePecker and more specifically
%studies the impact of control dependency in the code
%gadget, the imbalanced data processing, and the neural networks
%on the deep learning-based vulnerability detection. As
%discussed above, the extension is based on a completely new
%implementation using an extended open source tool Joern,
%because a straightforward extension cannot accommodate
%new semantic information for VulDeePecker which is based
%on the commercial tool Checkmarx.

%% file: sections/concl.tex
\section{Conclusion}
We present {\tool}, a novel DL-based approach to provide sub-graphs in
PDG, that explains the prediction results of graph-based vulnerability
detection.
%The following key ideas enable our approach (1) 
Our empirical evaluation on vulnerability databases shows that {\tool}
outperforms the existing DL-based approaches by 64\%--122\% and 105\%--255\%
in top-10 nDCG and MAP ranking scores.

Our key {\bf limitations} include 1) un-seen vulnerabilities, 2) the
vulnerable statements incorrectly identified due to data/control
dependencies with vulnerable ones, 3) missed vulnerable statements due
to multiple edges of data/control dependencies.

With {\tool} being a ML/DL-based vulnerability detection model, we aim to raise the level of ML/DL-based approaches, which are not able to point out the statements that caused the model to predict the vulnerability. Thus, we compared IVDetect with the detection approaches of the same category, rather than with static-analysis tools. In the future, we plan to compare {\tool} with static analysis tools.